\documentclass{optica-article}

\journal{opticajournal}

\articletype{Research Article}

\usepackage{upgreek}
\usepackage{graphicx}
\usepackage{subfigure}
\usepackage{booktabs}
\usepackage{indentfirst}
\usepackage{ragged2e}

\begin{document}

\title{Supercontinuum generation in dispersion engineered highly doped silica glass waveguides}

\author{Guangkuo Li,\authormark{1} Yuhua Li,\authormark{2} Feng Ye,\authormark{3} Qian Li,\authormark{3} Shao Hao Wang,\authormark{4} Benjamin Wetzel,\authormark{5} Roy Davidson,\authormark{6} Brent E. Little,\authormark{6} and Sai Tak Chu,\authormark{1,*}}

\address{\authormark{1}Department of Physics, City University of Hong Kong, Hong Kong, China\\
	\authormark{2}Key Laboratory of Optical Field Manipulation of Zhejiang Province, Department of Physics, Zhejiang Sci-Tech University, Hangzhou, China\\
	\authormark{3}School of Electronic and Computer Engineering, Peking University, Shenzhen, China\\
	\authormark{4}FZU-Jinjiang Joint Institute of Microelectronics, Jinjiang Science and Education Park, Fuzhou University, Jinjiang 362200, China\\
	\authormark{5}XLIM Research Institute, CNRS UMR 7252, University of Limoges, Limoges, France\\
	\authormark{6}QXP Technology Inc., Xi’an, China}

\email{\authormark{*}saitchu@cityu.edu.hk}

\begin{abstract*}
	\noindent We investigate the effect of a lower index oxide layer inclusion within a highly doped silica glass slot waveguide for optimized supercontinuum generation at telecom wavelengths. By controlling the thickness of the oxide slot, we demonstrate that one can engineer the waveguide dispersion profile so that to obtain supercontinua with vastly different spectral broadening dynamics and bandwidths. Using this approach, we designed and fabricated a waveguide with a low and flat dispersion profile of less than 43 $\rm {ps/km/nm}$ across a wavelength range spanning over 1000 nm. We show that, when pumped at the telecom C-band, we can generate a supercontinuum that spans over 1.5 octave, from approximately 817 nm to 2183 nm. The numerical simulations, whose parameters are derived from the measured waveguide dimension and material indices, exhibit good agreement with experimental measurements, where one can observe both a qualitative and quantitative match in the supercontinuum overall spectrum and specific features (e.g. soliton and dispersive wave locations). This study represents an important step forward in the control and manipulation of dispersive and nonlinear dynamics in highly doped silica glass waveguides, paving the way towards advanced on-chip broadband light manipulation.
\end{abstract*}

\section{Introduction}

\noindent Supercontinuum (SC) generation arises from the propagation of a high-power optical pump through a nonlinear medium. This highly complex dynamical process involves both the nonlinearity and dispersion of the medium, and can be extremely sensitive to these characteristics as well as those of the input pump signal. Due to the variety of SC broadening scenario and associated output spectral properties \cite{sylvestre2021recent}, SC have found numerous applications as ultra-broadband light sources in many fields such as frequency metrology \cite{udem2002optical,lamb2018optical}, spectroscopy \cite{kano2003characterization,amiot2017cavity}, and optical coherence tomography \cite{hartl2001ultrahigh,froehly2012supercontinuum}. In addition, SC broadening dynamics have also been investigated from a more fundamental viewpoint, being a testbed for the study of exotic phenomena such as rogue wave \cite{solli2007optical,dudley2008harnessing} and dark soliton \cite{milian2017spectral,zhao2019generation}, or even exploited as an onset nonlinear broadening stage in cascaded architectures towards ultrashort UV or mid-IR pulse formation \cite{sylvestre2021recent}.

The introduction of photonic crystal fibers (PCF) has made these light sources more accessible as the group velocity dispersion (GVD) of PCFs can be engineered to enable strong nonlinear interactions over significant fiber propagation lengths \cite{dudley2006supercontinuum}. Such dispersion engineering lower the energy requirement for the nonlinear spectral broadening process, allowing the replacement of bulky high power OPO laser by cost-effective semiconductor diode lasers to serve as the optical pump \cite{cordeiro2005engineering,wadsworth2000soliton}. Besides optical fibers, SC can also be generated in integrated optical waveguides (playing the role of the nonlinear propagation medium), which have been demonstrated in platforms spanning silicon \cite{lau2014octave}, silicon nitride \cite{johnson2015octave}, silicon germanium \cite{sinobad2018mid}, chalcogenide \cite{tremblay2018picojoule}, aluminum nitride \cite{lu2020ultraviolet}, InGaP-on-insulator \cite{dave2015dispersive} and highly doped silica glass (HDSG) \cite{duchesne2010supercontinuum}. Interestingly, these waveguides generally exhibit both higher nonlinearity and tighter mode confinement compared to optical fibers. The key aspect of generating the SC spectrum in an integrated fashion is to allow for the direct connection of the broadband light source with other waveguide components on the same chip. The advantage is thus two-fold: integrated approaches not only drastically reduce the overall footprint of the device (and overall system), they can also provide direct access to a host of optical functionalities compatible with advanced and reconfigurable optical pulse processing within a single (stable) on-chip photonic framework \cite{wetzel2018customizing,bogaerts2020programmable}.

Among those integrated architectures, HDSG constitutes a versatile CMOS-compatible platform that possesses very low linear and nonlinear losses. The Kerr nonlinearity of $n_2 = 1.15\times 10^{-19}\ {\rm m^2/W}$ with nonlinear coefficient $\gamma$ of approximately two hundred times larger than the single mode fibers (SMF). Its high nonlinear figure-of-merit \cite{moss2013new} makes the platform highly efficient in nonlinear applications \cite{ferrera2008low,razzari2010cmos,sun2019integrating,xu202111}. In 2010, Duchesne et. al., demonstrated SC generation in the HDSG spiral waveguide using a 45 cm long channel waveguide with core dimension of 1.45 $\rm {\upmu m}$ by 1.50 $\rm {\upmu m}$ and with a refractive index of $n = 1.7$ \cite{duchesne2010supercontinuum}. Using a 100 fs pulse with 1450 W peak power to pump the waveguide anomalous dispersion region at 1550 nm, they obtained a SC spectral bandwidth of 300 nm. However, the SC spectrum coverage was limited by the dispersion profile of the waveguide. In fact, the span of the obtained SC spectrum is comparable to only the anomalous dispersion region of the waveguide. To boost SC spectral broadening dynamics, it is generally desirable to design waveguide dispersion profiles that are flat across a wide spectral range, and ideally featured with low anomalous dispersion: the flat and low anomalous GVD region indeed limits the temporal walk-off (and eventual pulse spreading) during SC formation, thus maximizing the phase matching and enhance the nonlinear spectral broadening throughout propagation \cite{lafforgue2022supercontinuum}. Furthermore, such dispersion engineering approach can also be leveraged to adjust the zero-dispersion wavelength (ZDW) location and therefore tailor the waveguide for an optimized pump wavelength \cite{guo2018mid}.

In this work, we report a systematic study of the role and impact of a thin oxide layer inclusion within the core of HDSG channel waveguides for SC generation. In particular, we analyze how the thickness of the oxide layer can be used to engineer the dispersion profile of the waveguide and the corresponding output SC spectrum. We show that using a 0.1 $\rm {\upmu m}$ oxide slot extends the SC spectrum coverage to over 1.5 octaves, which is 664 nm wider compared to SC generated from the conventional channel waveguide without the slot. We also perform numerical simulations based on a generalized nonlinear Schrödinger equation (GNLSE) model to study soliton dynamics, spectral broadening processes and SC generation in the fabricated HDSG waveguides. We show that numerical results are in good agreement with experimental measurements, and that the generated supercontinuum exhibits both phase and intensity stability attested by a high coherence degree and a high signal-to-noise ratio (SNR) across its entire spectral bandwidth.

\section{Dispersion Engineering}

\noindent Dispersion plays a critical role in the spectral broadening of optical pump signals in a nonlinear medium. For instance, soliton existence and propagation physically require anomalous GVD waveguides to produce a negative chirp able to balance the positive chirp produced by self-phase modulation (SPM). In contrast, in the normal GVD region, the pulse is stretched due to the uncompensated SPM positive chirp, which in turn limits its peak power and associated SPM efficiency. Thus, a wide anomalous GVD region with a flat and close-to-zero dispersion can reduce the walk-off between the high and low frequency components and strengthen the nonlinear effects and further gradually compresses the pulse width and broaden the spectral span of SC correspondingly. A popular structure to allow flexible engineering of the waveguide dispersion is the slot waveguide structure. It has a thin layer of lower index oxide, the slot, placed in the waveguide core. This type of structure was shown effective in engineering the dispersion profile of the silicon rich waveguides and greatly expanded their SC spectrum \cite{zhang2011chip,zhu2012ultrabroadband}. The effectiveness of the slot structure will be investigated in the dispersion engineering of the HDSG waveguides in this study.

\begin{figure}[htbp]
	\centering
	\subfigure[]{
		\raisebox{0pt}{\includegraphics[width = 0.200\textwidth]{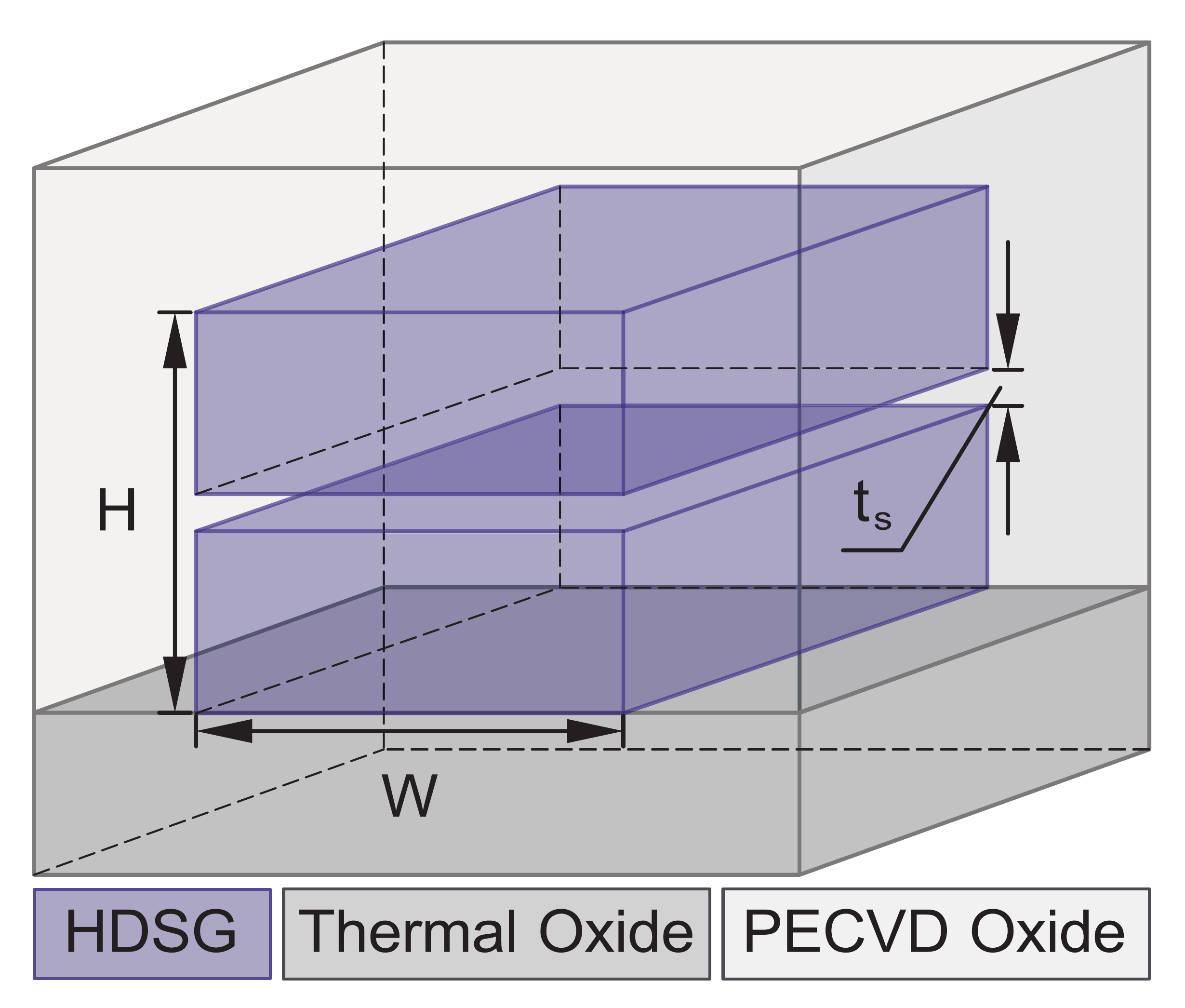}}
	}
	\subfigure[]{
		\raisebox{2pt}{\includegraphics[width = 0.203\textwidth]{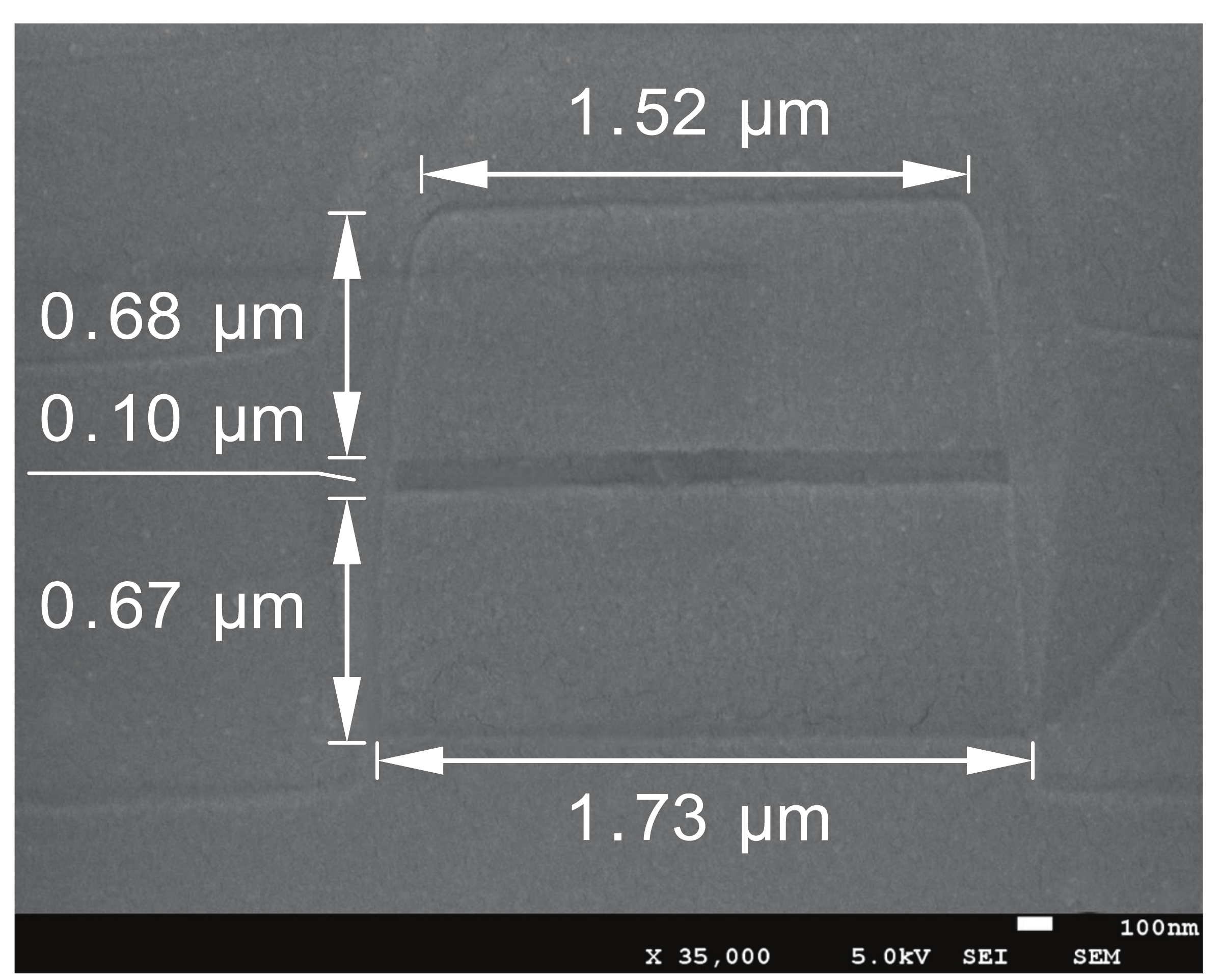}}
	}
	\subfigure[]{
		\raisebox{-5pt}{\includegraphics[width = 0.235\textwidth]{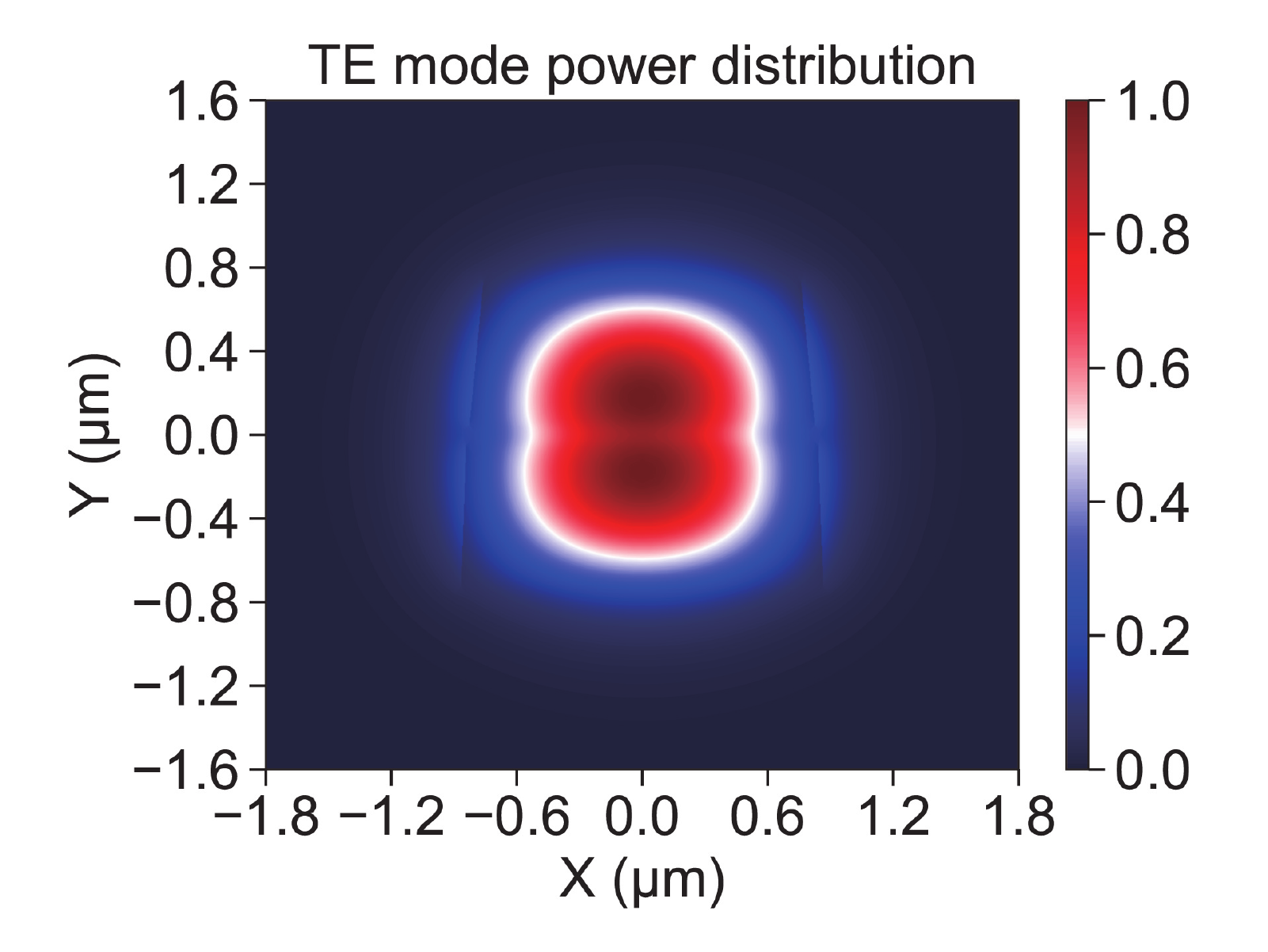}}
	}
	\subfigure[]{
		\raisebox{-5pt}{\includegraphics[width = 0.235\textwidth]{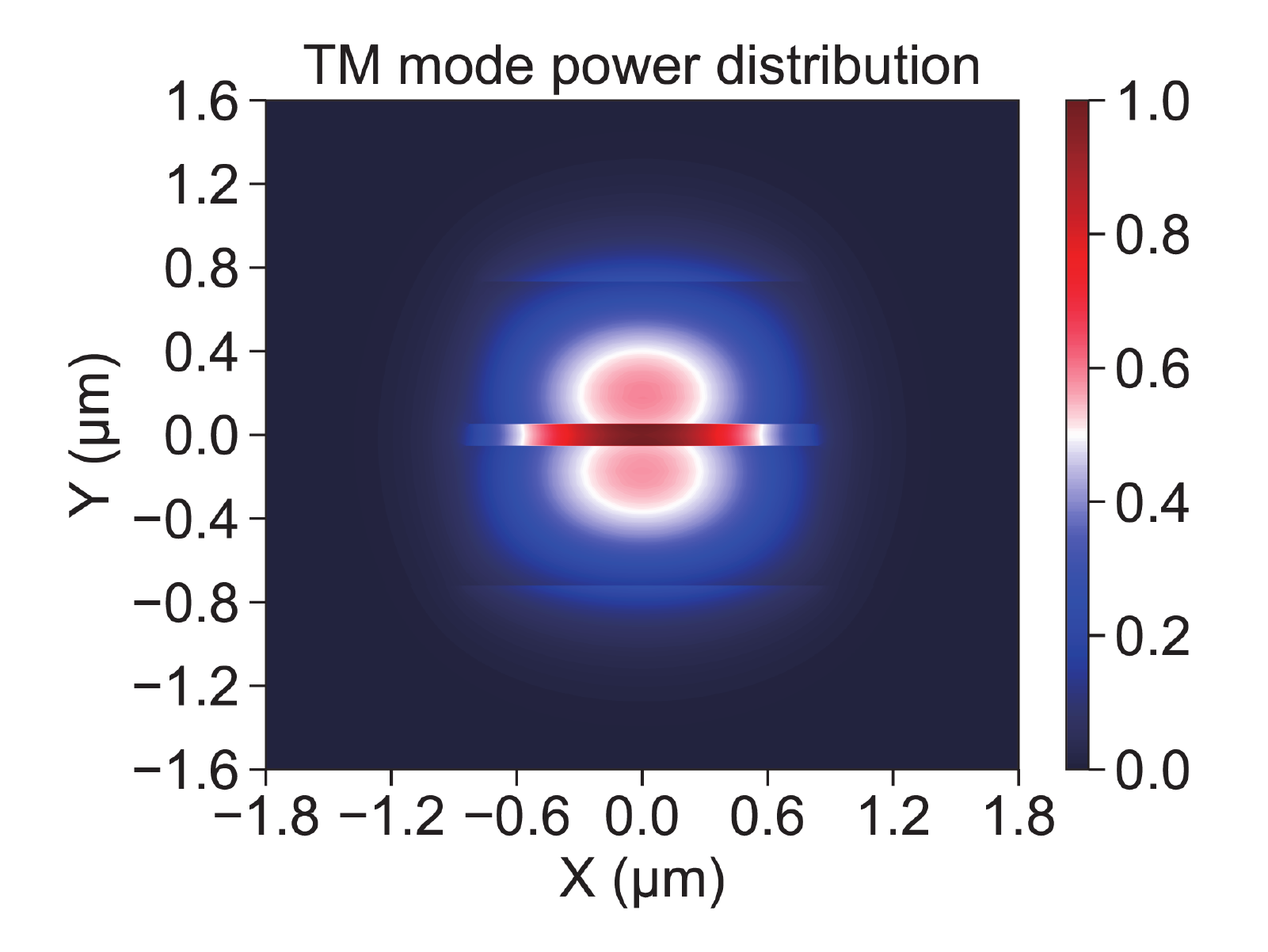}}
	}
	\caption{(a) Cross-section of the selected waveguide structure. (b) SEM image of the fabricated waveguide. (c) Normalized TE mode power distribution. (d) Normalized TM mode power distribution.}
\end{figure}

The schematic of the slot waveguide selected for this study is shown in Fig. 1(a), having the oxide slot placed at the center of the $n = 1.7$ HDSG core. Except for the thermal oxide lower cladding, the rest of the waveguide structure was fabricated with plasma-enhanced chemical vapor deposition (PECVD). Since the indices of the thermal and PECVD oxides are slightly difference, both the thermal and PECVD oxide indices were used in the calculation of the effective index of the waveguides. Figure 1(b) shows the scanning electron microscope (SEM) image of the cross-section of the fabricated slot thickness $\rm {t_s=0.1\ \upmu m}$ waveguide. The SEM image indicates that the fabricated waveguide core has a composite dimension of 1.73 $\rm {\upmu m}$ by 1.45 $\rm {\upmu m}$, with a slight angle of approximately 4.1 degrees. The normalized power distribution for both the fundamental TE and TM modes are shown in Figs. 1(c) and 1(d). For TE polarization, the slot splits the power into the upper and lower core regions, where most of the energy is confined. In contrast, the bulk of the TM mode power lies in the vicinity of the slot region due to the discontinuity of the E-field at the slot interfaces.

\begin{figure}[htbp]
	\centering
	\subfigure[]{
		\includegraphics[width = 0.31\textwidth]{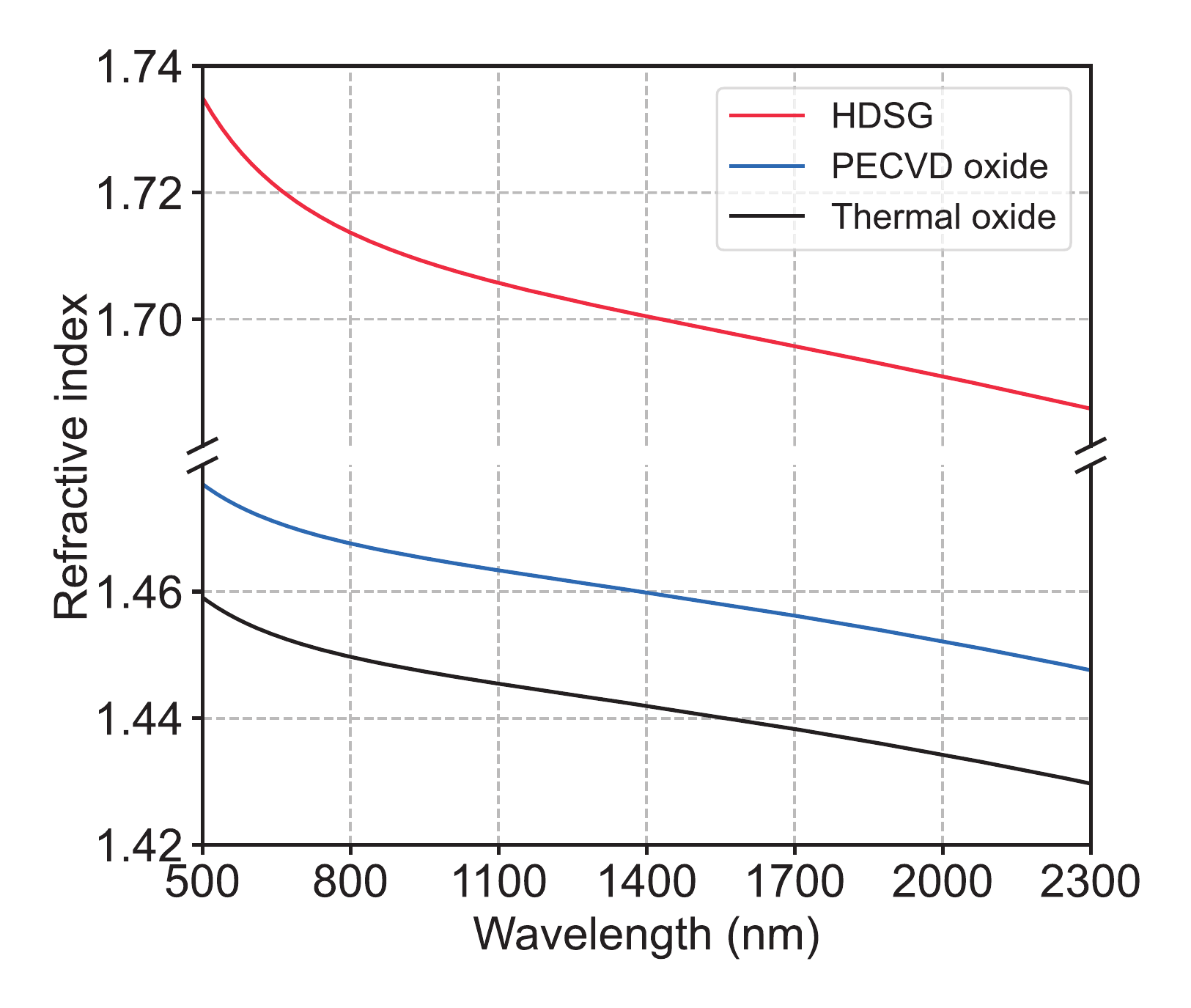}
	}
	\subfigure[]{
		\includegraphics[width = 0.31\textwidth]{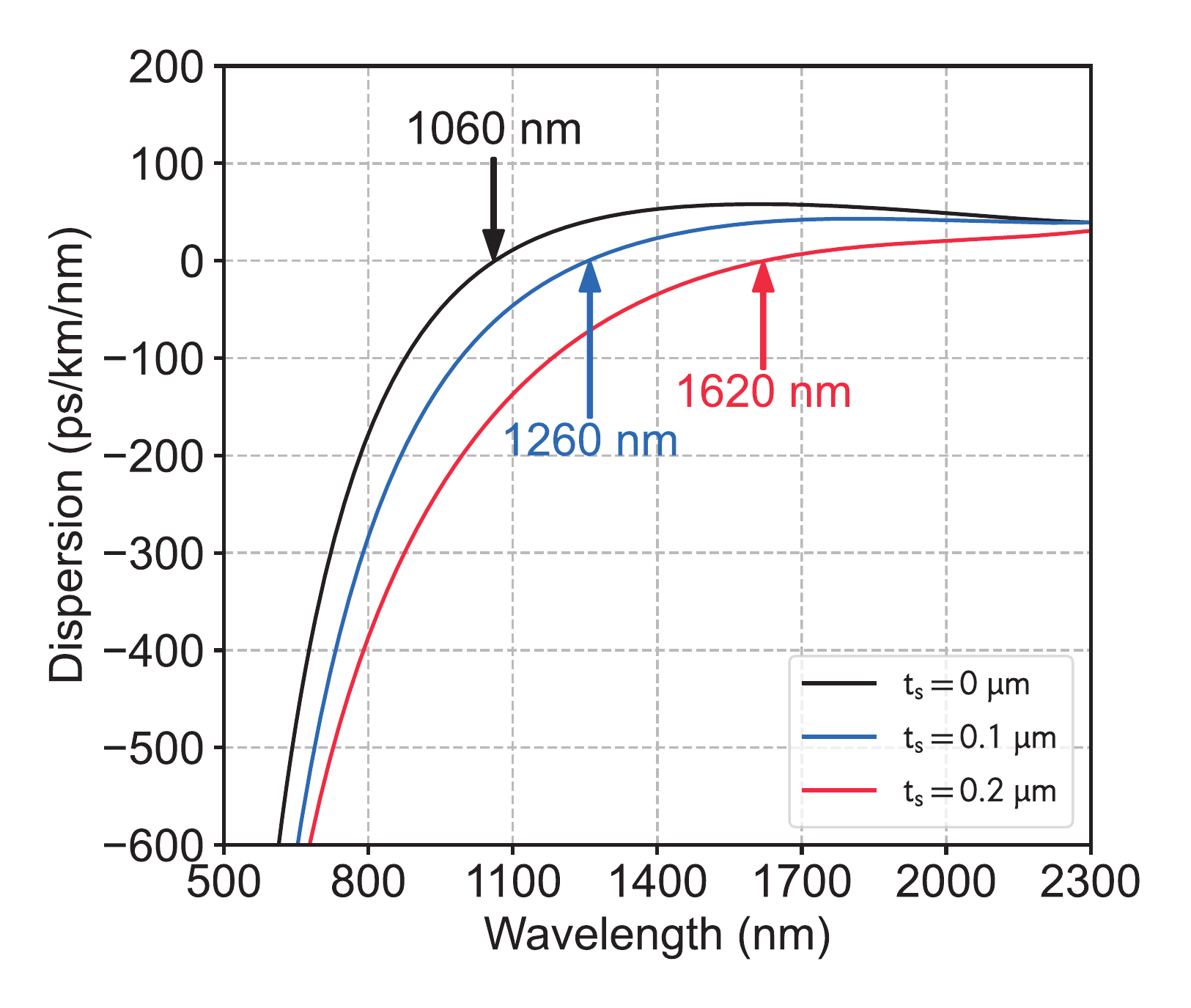}
	}
	\subfigure[]{
		\includegraphics[width = 0.31\textwidth]{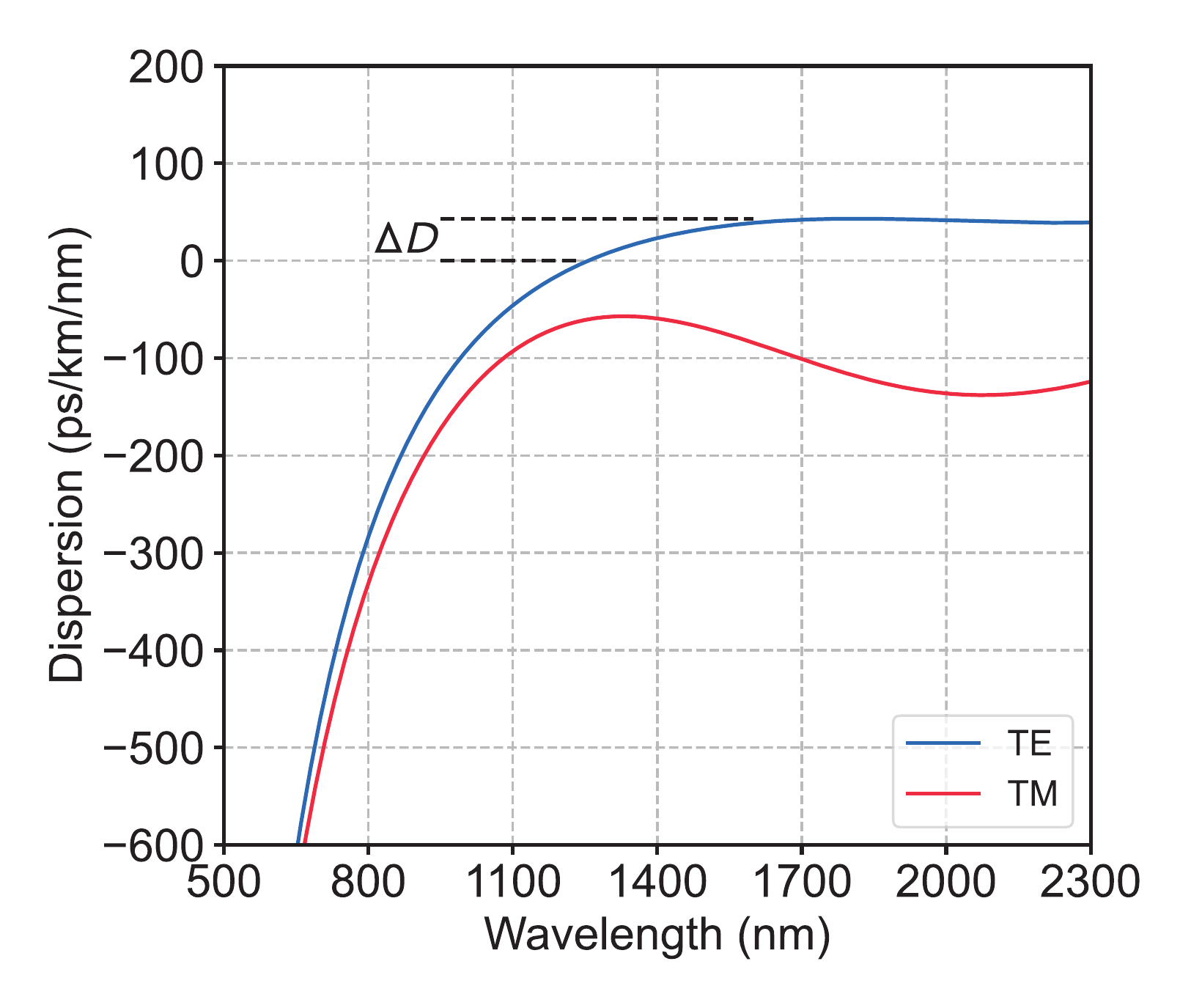}
	}
	\caption{(a) Refractive index for thermal oxide (black), PECVD oxide (blue) and HDSG (red). (b) TE mode dispersion profiles in $\rm {t_s=0,\ 0.1\ \rm {and}\ 0.2\ \upmu m}$ waveguides. (c) TE (blue) and TM (red) modes dispersion profiles of $\rm {t_s=0.1\ \upmu m}$ waveguide.}
\end{figure}

The material indices used to calculate the waveguide dispersion are shown in Fig. 2(a), where the indices from 500 nm to 1690 nm were measured with an ellipsometer while the indices from 1690 nm to 2300 nm were extended by the Sellmeier equation (based on the index measurements). Figure 2(b) shows the simulated TE mode dispersion profile of a 1.8 $\rm {\upmu m}$ by 1.5 $\rm {\upmu m}$ waveguide for different slot thickness $\rm {t_s}$ between 0 to 0.2 $\rm {\upmu m}$. It shows that an increase of the slot thickness induces a downward shift to the dispersion curve and moves the ZDW location toward the longer wavelength end. In these cases, the ZDW is shifted from 1060 nm to 1620 nm by the addition of a 0.2 $\rm {\upmu m}$ oxide slot. The dispersion curves of both the TE and TM modes for a waveguide with $\rm {t_s=0.1\ \upmu m}$ are displayed in Fig. 2(c). The figure shows that the TM mode is always in the normal dispersion regime. Conversely, the TE mode has a very flat and low anomalous dispersion spanning from 1260 nm to over 2300 nm, with a $\Delta D$ lower than 43 $\rm {ps/km/nm}$ across the whole region.

\section{Results and Discussions}

\subsection*{Supercontinuum measurements}

\noindent The pump source used in the experiment was a femtosecond pulse fiber laser (Amonics AFL-1560-FS- 300-03-AOCP-SA) centered at 1560 nm with full width at half maximum (FWHM) pulse duration adjustable from 180 fs to 293 fs, and a corresponding to pulse energy $E_{\rm p}$ ranging from 621 pJ to 280 pJ at repetition rate of 40 MHz. The pulse duratuion was retrieved from GRENOUILLE measurements (Model 15-100-USB) \cite{trebino2000frequency}. Separate optical spectrum analyzers (OSA) were used to record the SC spectra in the experiment (YOKOGAWA AQ6373B for 500 to 800 nm, AQ6370D for 800 to 1600 nm and AQ6375 for 1600 to 2300 nm).

\begin{figure}[htbp]
	\centering
	\subfigure[]{
		\includegraphics[width = 0.316\textwidth]{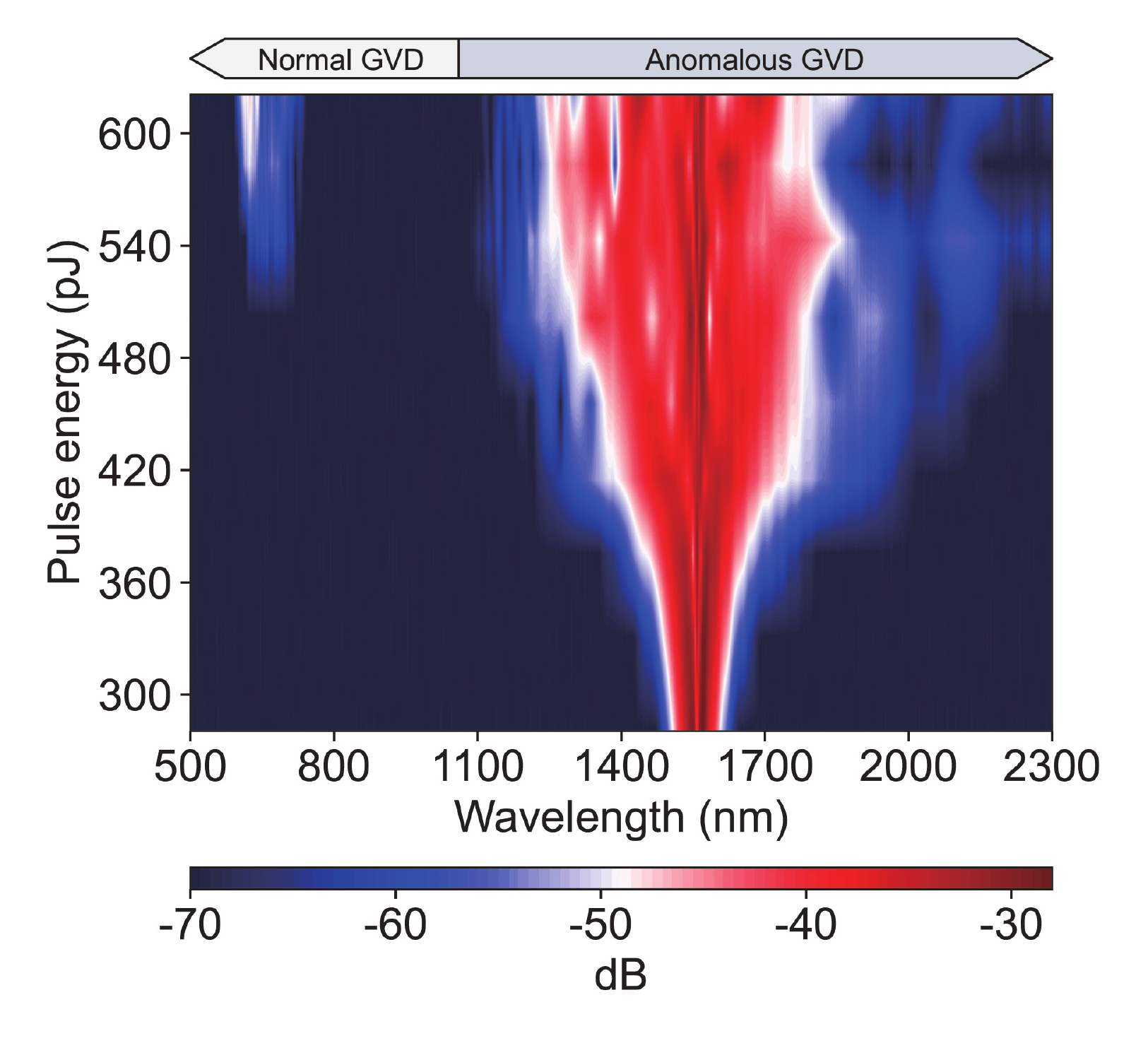}
	}
	\subfigure[]{
		\includegraphics[width = 0.316\textwidth]{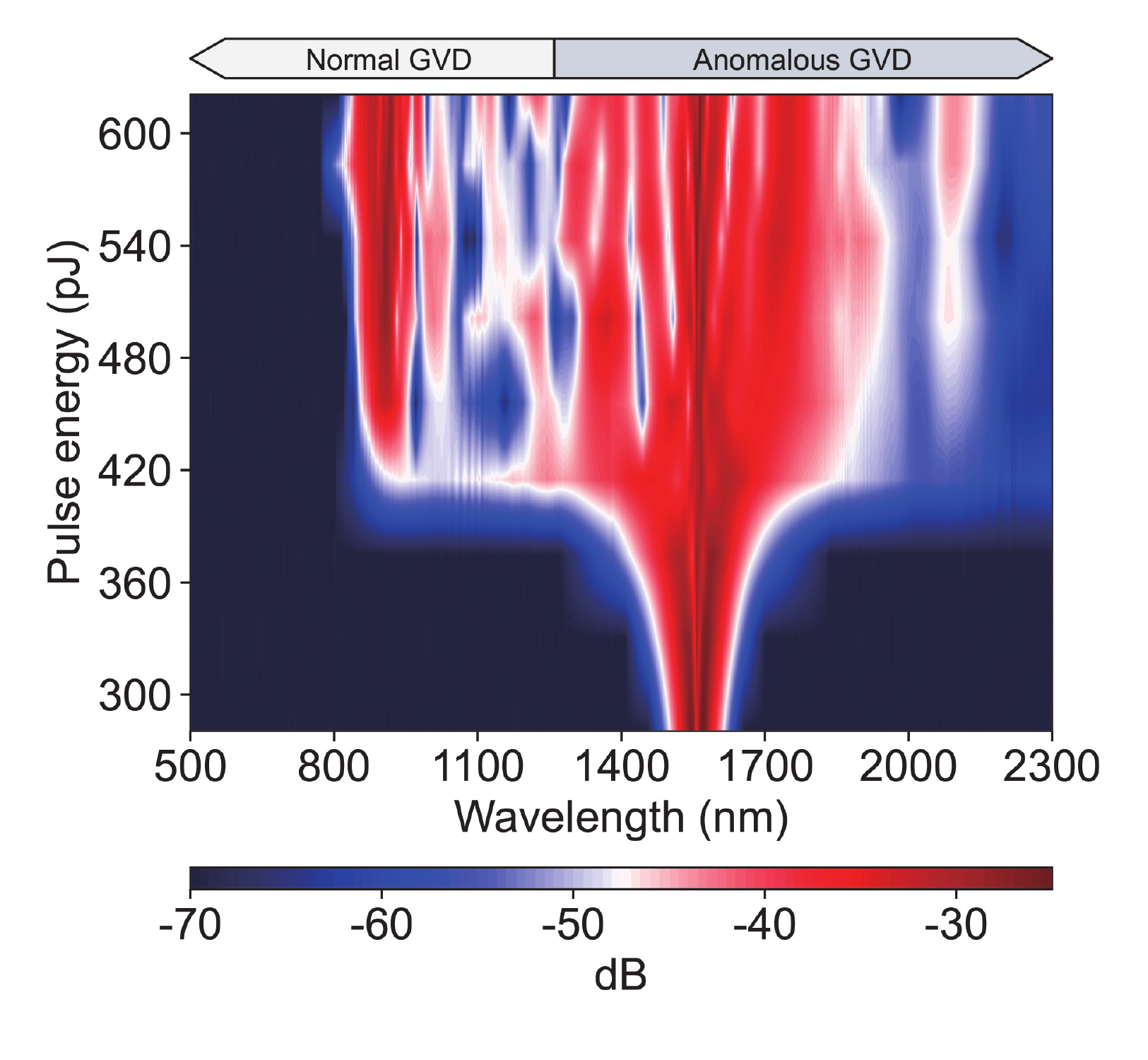}
	}
	\subfigure[]{
		\includegraphics[width = 0.316\textwidth]{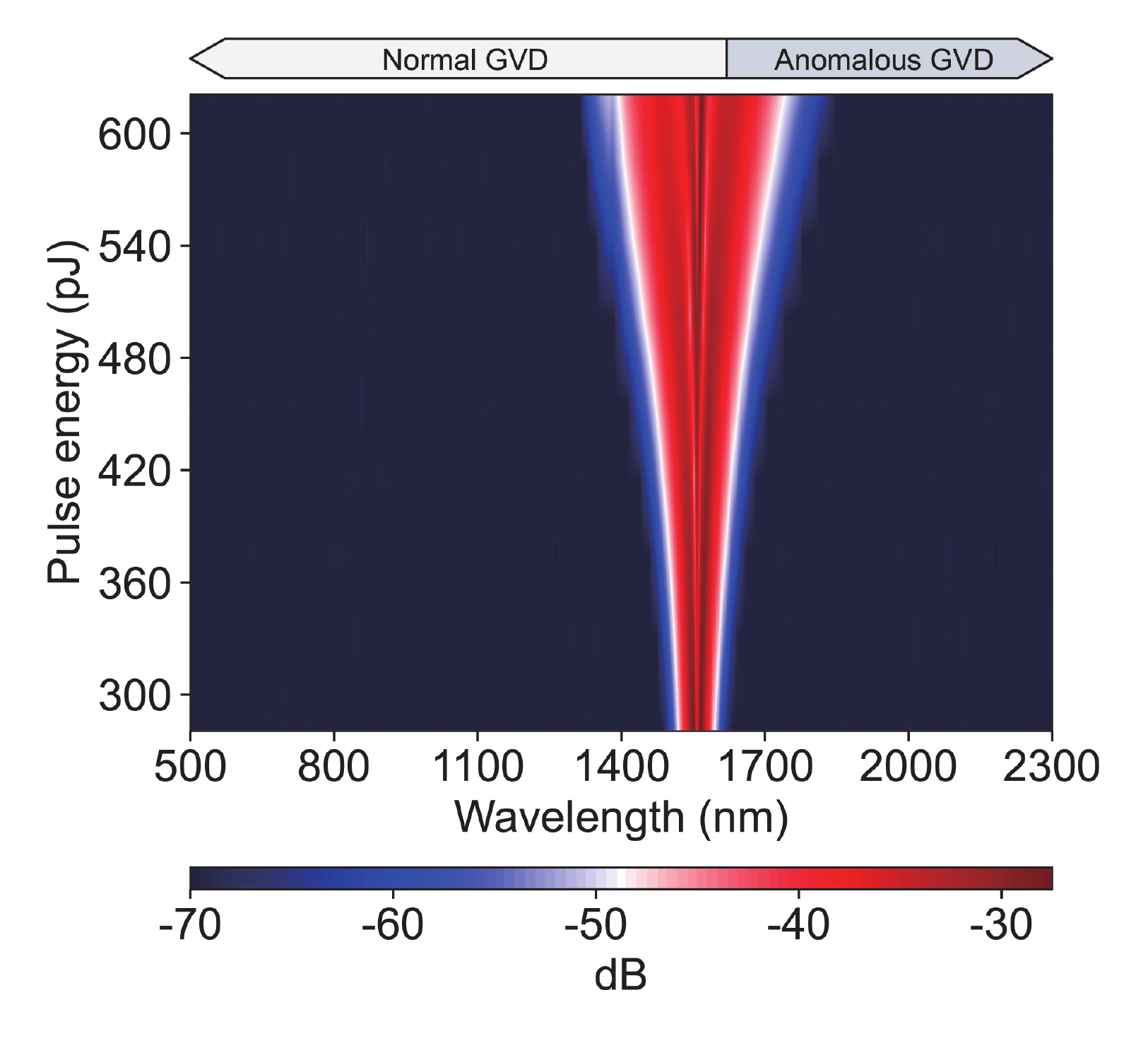}
	}
	\subfigure[]{
		\includegraphics[width = 0.31\textwidth]{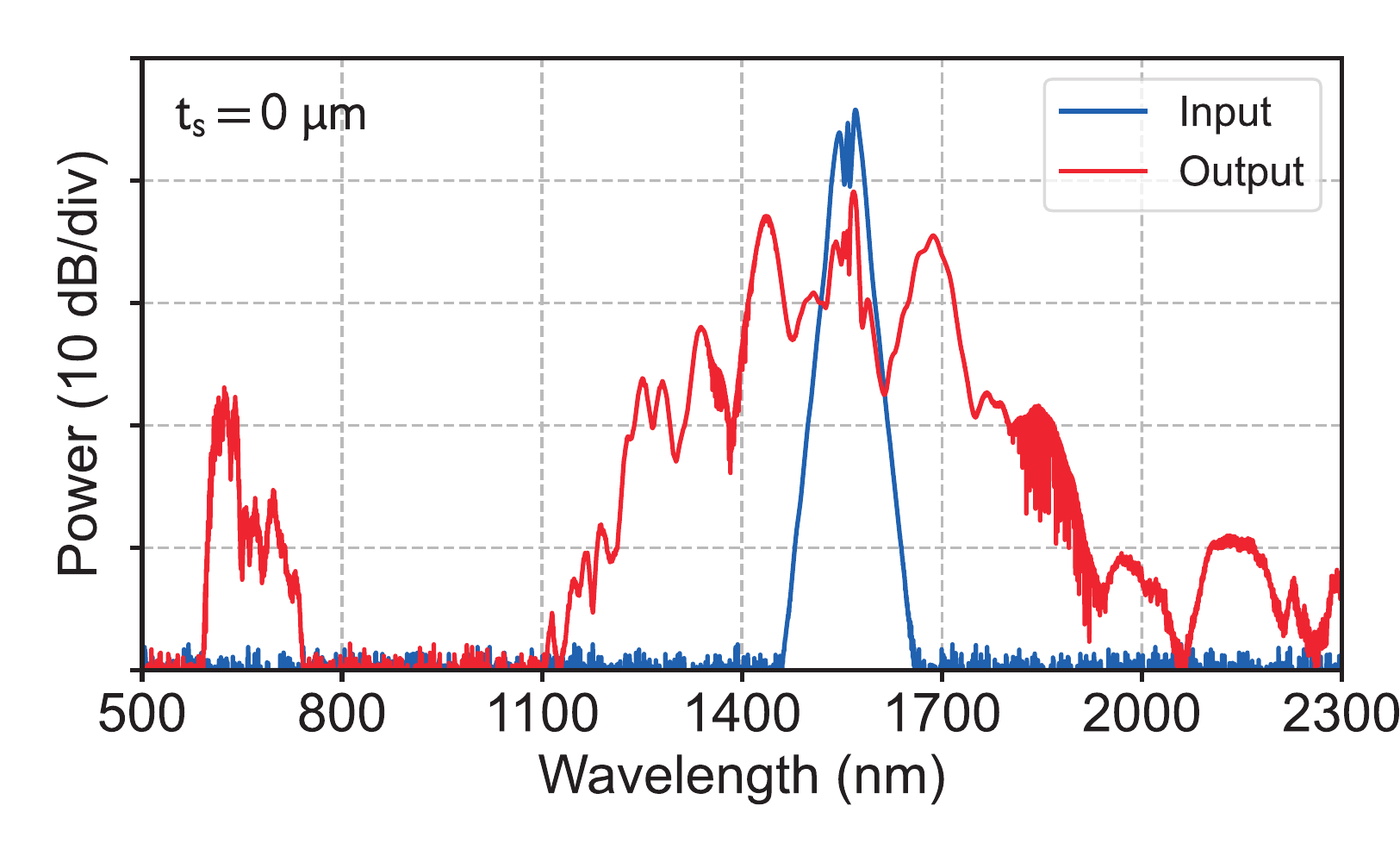}
	}
	\subfigure[]{
		\includegraphics[width = 0.31\textwidth]{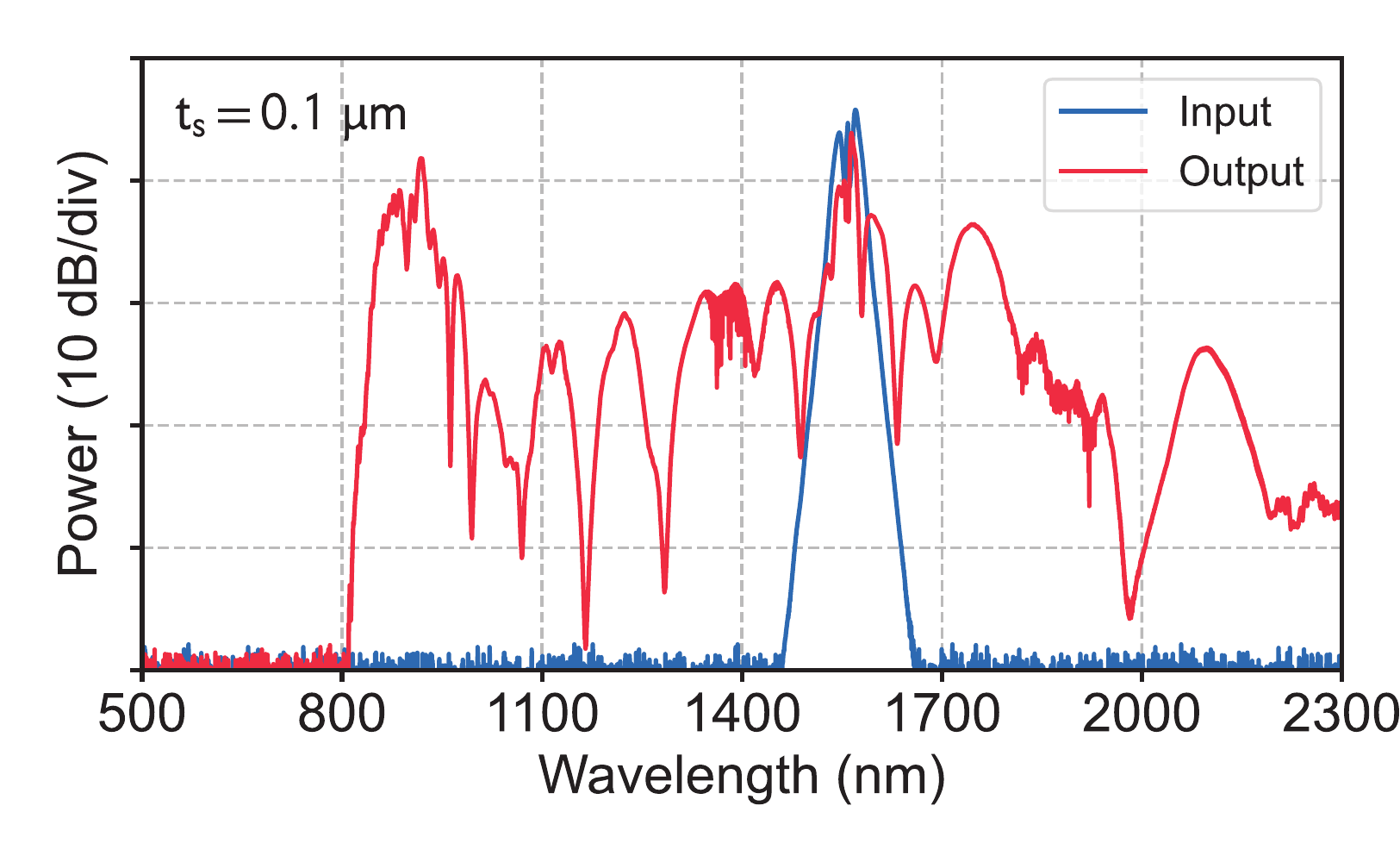}
	}
	\subfigure[]{
		\includegraphics[width = 0.31\textwidth]{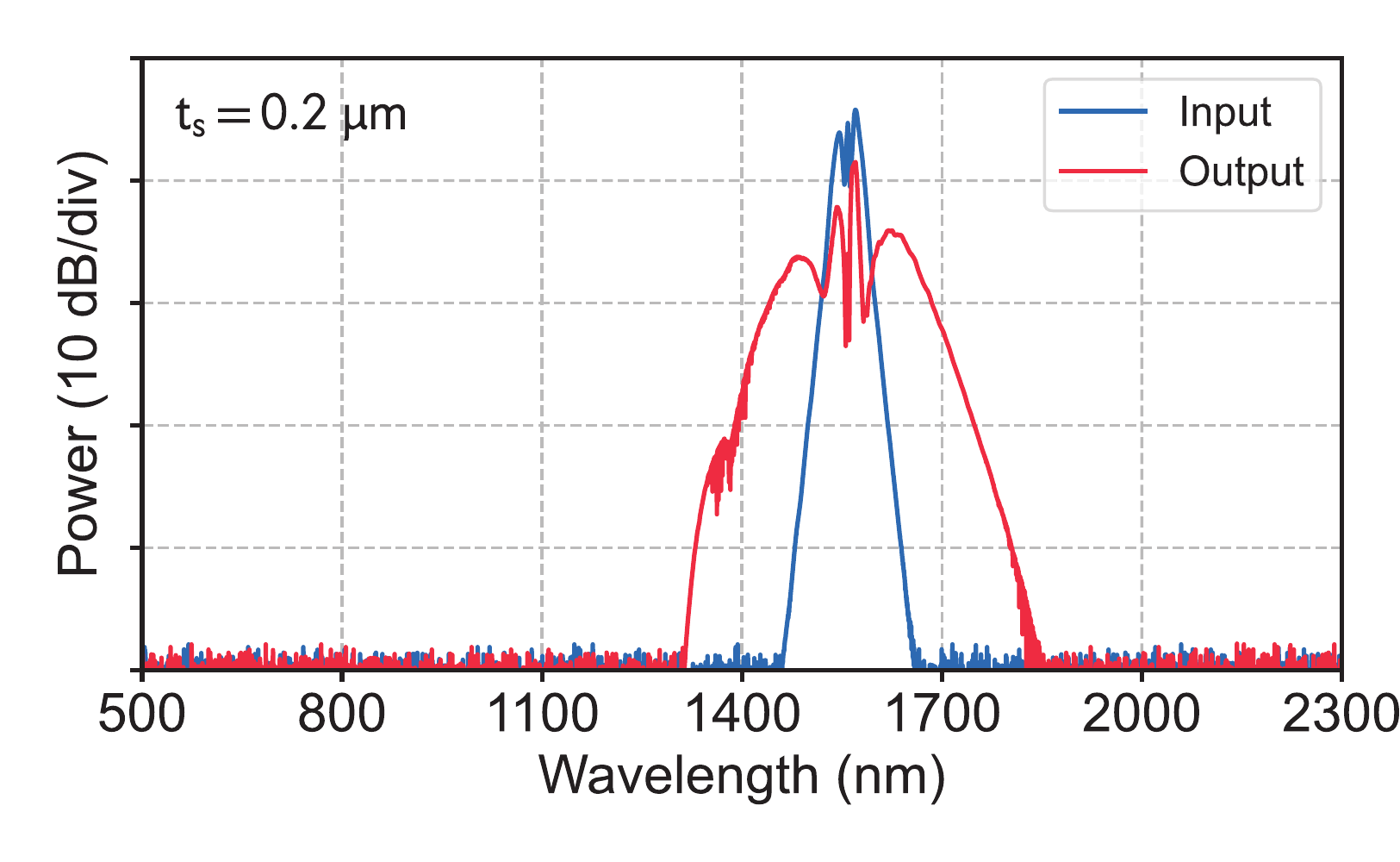}
	}
	\caption{Comparison of measured SC generation in $\rm {t_s=0,\ 0.1\ \rm {and}\ 0.2\ \upmu m}$ waveguides: (a-c) SC evolution vs input pulse energy; (d-f) output spectrum at $E_{\rm p}=621\ {\rm pJ}$, corresponds to pulse width of 180 fs and peak power of 2835 W propagating through 3.1 cm long waveguides.}
\end{figure}

Figures 3(a-c) show the measured spectral evolution within three waveguides having the same length of 3.1 cm but featured with different dispersion profiles. At 1560 nm, the pump wavelength is in the anomalous GVD region for the waveguides with $\rm {t_s=0\ {\rm and}\ 0.1\ \upmu m}$, and in the normal GVD region for the $\rm {t_s=0.2\ \upmu m}$ waveguide. When pumped in anomalous GVD region, as seen in Figs. 3(a) and 3(b), the spectra experience a similar slow broadening process under the effect of SPM when the energy $E_{\rm p}$ is below 400 pJ. Although both spectra experience a stronger broadening for $E_{\rm p} > 400\ {\rm pJ}$, the spectrum for $\rm {t_s=0.1\ \upmu m}$ spreads at a much faster rate than the spectrum for $\rm {t_s=0\ \upmu m}$ due to its lower dispersion. In fact, at $E_{\rm p} = 450 \ {\rm pJ}$, the spectrum for the waveguide with a slot of 0.1 $\rm {\upmu m}$ had spread over 1200 nm, close to the maximum bandwidth reached in this experiment. In comparison, for the waveguide without slot, the spectrum had expended to approximately 500 nm and continue to spread for higher energy values of $E_{\rm p}$.

For these two cases with the pump in the anomalous GVD region, soliton fission occurs when $E_{\rm p}$ is between 380 pJ to 420 pJ, such high order soliton dynamics play a key role in the expansion of the spectra after the initial SPM stage. When the pulse is compressed to its Fourier limit under the combined action of GVD and SPM, higher-order dispersion and symmetry breaking effects (e.g. nonlinear self-steepening) perturbs the pulse, inducing soliton fission where individual solitons are ejected from the initial pulse and are shifted towards longer wavelengths via soliton self-frequency-shift (SSFS) \cite{lee2008soliton}. Here, each ultrashort fundamental or low-order soliton carries a large number of different frequency components to form a broadband SC spectrum. Conversely, through their spectral shift and associated reshaping \cite{agrawal2019nonlinear}, the ejected solitons shed part of their energy to a linear dispersive wave (DW) in the normal GVD region, where the phase matches with the soliton in anomalous GVD region. For the $\rm {t_s=0\ \upmu m}$ waveguide, because of the large distance between the pump and the normal GVD region, the generated SC spectrum is mostly confined within the anomalous region, where the DW only starts to appear when $E_{\rm p}$ reaches a relatively high value of 500 pJ (and is detached from the main body of the spectrum). In contrast, the addition of the slot flattens the dispersion curve of the $\rm {t_s=0.1\ \upmu m}$ waveguide allowing the SC spectrum to spread across the anomalous region into the normal region to generate a large DW peak. In this case, DW generation begins shortly after soliton fission and the resulting DW peak is connected to the central part of the SC spectrum. As for the broadening of the $\rm {t_s=0.2\ \upmu m}$ waveguide, illustrated in Fig. 3(c), it is mainly due to the SPM and is much narrower than the previous two cases. With the pump in the normal GVD region, pulse propagation cannot induce any high order soliton dynamics nor satisfying the required phase matching condition, so that no soliton fission nor DW generation is triggered.

Figures 3(d-f) show the input and output spectra for the three waveguides pumped at $E_{\rm p} = 621 \ {\rm pJ}$, with 180 fs FWHM pulse duration and 2835 W peak power. In Fig. 3(a), the spectrum of the $\rm {t_s=0\ \upmu m}$ waveguide with 30 dB bandwidth spans from 1210 nm to 1912 nm with the DW located at 620 nm. While the dispersion engineered $\rm {t_s=0.1\ \upmu m}$ waveguide exhibits 664 nm wider bandwidth shown in Fig. 3(b), spanning from 817 nm to 2183 nm which includes the DW at 900 nm. For the $\rm {t_s=0.2\ \upmu m}$ waveguide, shown in Fig. 3(c), it produces a classical symmetrical spectrum between 1335 and 1796 nm that is due to SPM broadening.

\subsection*{Supercontinuum simulations}

\noindent Numerical simulation based on the GNLSE model \cite{dudley2006supercontinuum} were performed to investigate the spectro-temporal behavior of the SC generation process in these waveguides. Both the GVD and higher-order dispersion are accounted for in the numerical model. We determined the nonlinear parameters of the waveguides using the rigorous definition from \cite{sato2015rigorous}. It yields $\gamma=0.170,\ 0.148\ {\rm and}\ 0.135\ {\rm /W/m}$ for the TE mode of the $\rm {t_s=0,\ 0.1\ \rm {and}\ 0.2\ \upmu m}$ waveguides, respectively. This shows that the introduction of the oxide slot has slightly reduced the waveguide nonlinearity. Figures 4(a-c) show the simulated SC spectra using the input pulse shape retrieved from the experimental measurement of Figs. 3(a-c). Except for the spectral region above 2000 nm, the simulated results agree well with the measurements and are able to replicate most of the features observed in the experimental spectra, including the locations of the DW for the $\rm {t_s=0\ \rm {and}\ 0.1\ \upmu m}$ waveguides. The discrepancies in the longer wavelength region is mainly due to HDSG possessing a strong absorption peak at 2030 nm, which leads to the appearance of artifacts ("peaks" at 2100 nm) in both $\rm {t_s=0\ \rm {and}\ 0.1\ \upmu m}$ waveguides. The measured DW positions of these waveguides coincide very well with the phase mismatch $\beta_{\rm int} = 0$ shown in the upper panels of Figs. 4(a) and 4(b):

\begin{equation}
	\beta(\omega)-\beta(\omega_{\rm s}) - (\omega - \omega_{\rm s})v_g^{-1} = \sum\limits_{{\rm m} \ge 2} \frac{(\omega-\omega_{\rm s})^{\rm m}}{{\rm m}!}\frac{{\partial ^{\rm m}}}{\partial {\omega^{\rm m}}}\beta(\omega_{\rm s}) = \beta_{\rm int},
\end{equation}

\noindent where $\beta(\omega)$ is the propagation constant, $\omega_{\rm s}$ is the soliton center frequency, and $v_g$ is the group velocity \cite{guo2018mid}. We note that the addition of a lower index slot has shifted the phase matching position toward a longer wavelength, thus easing the DW generation process for $\rm {t_s=0.1\ \upmu m}$ waveguide.

\begin{figure}[htbp]
	\centering
	\subfigure[]{
		\includegraphics[width = 0.31\textwidth]{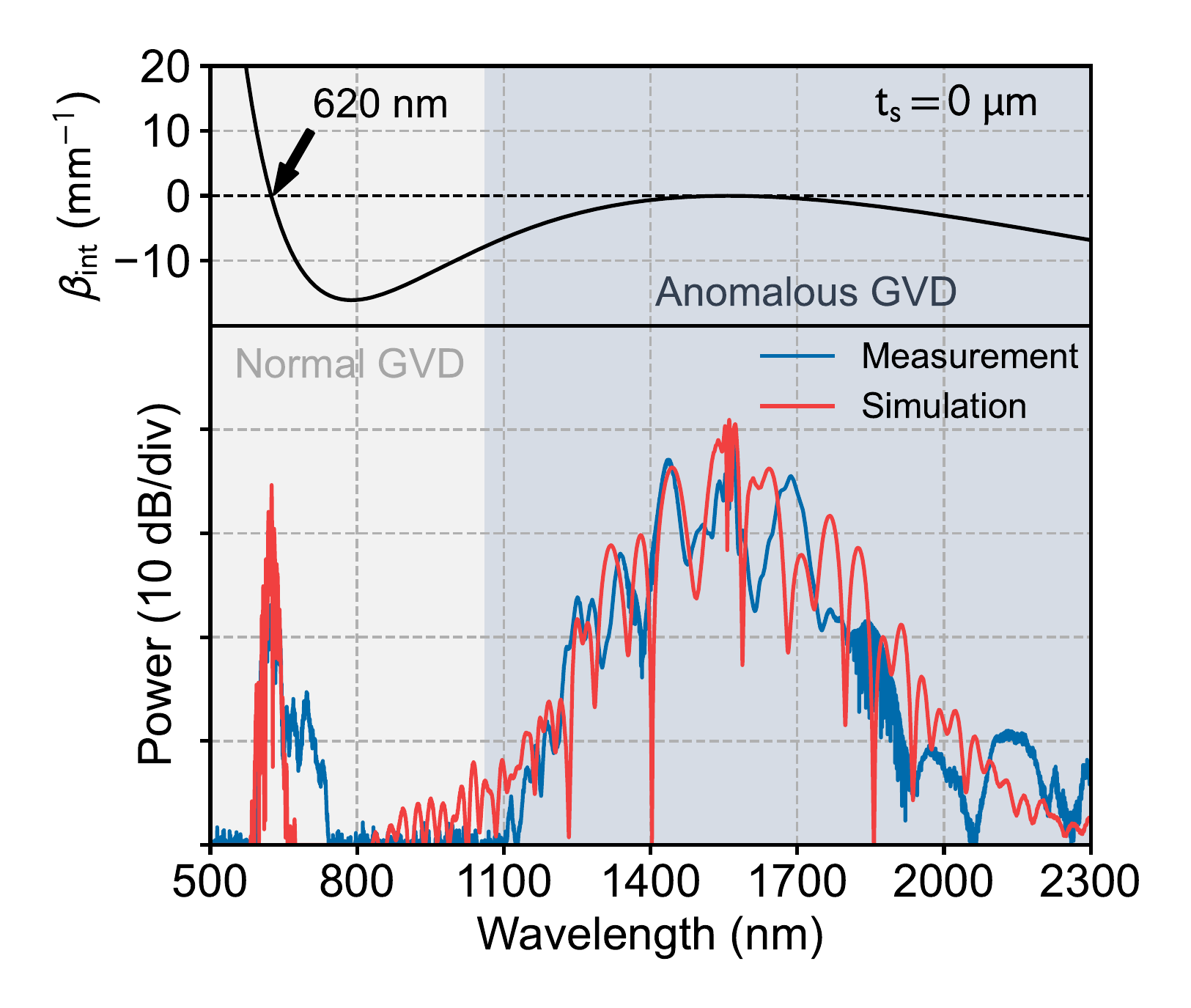}
	}
	\subfigure[]{
		\includegraphics[width = 0.31\textwidth]{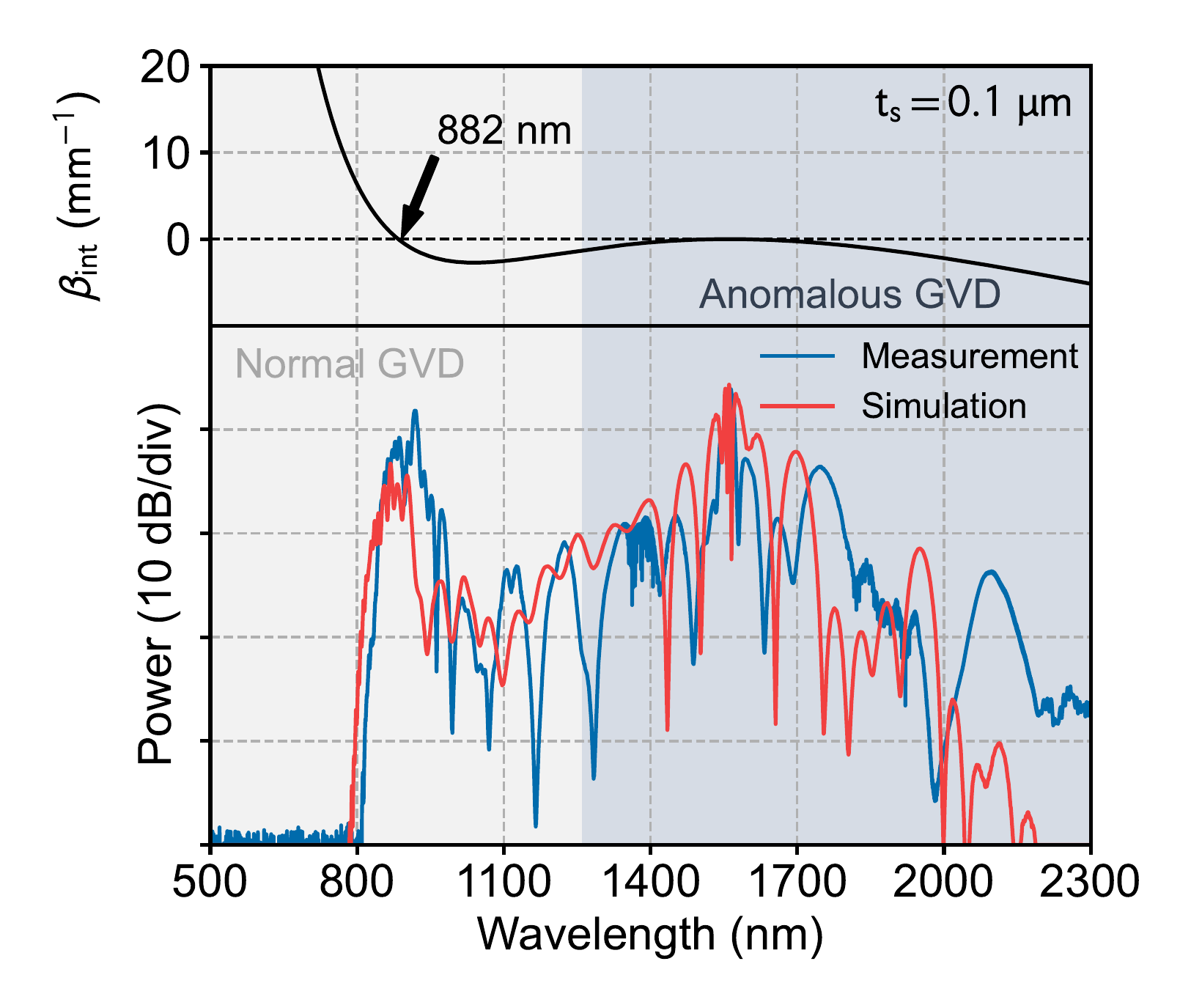}
	}
	\subfigure[]{
		\includegraphics[width = 0.31\textwidth]{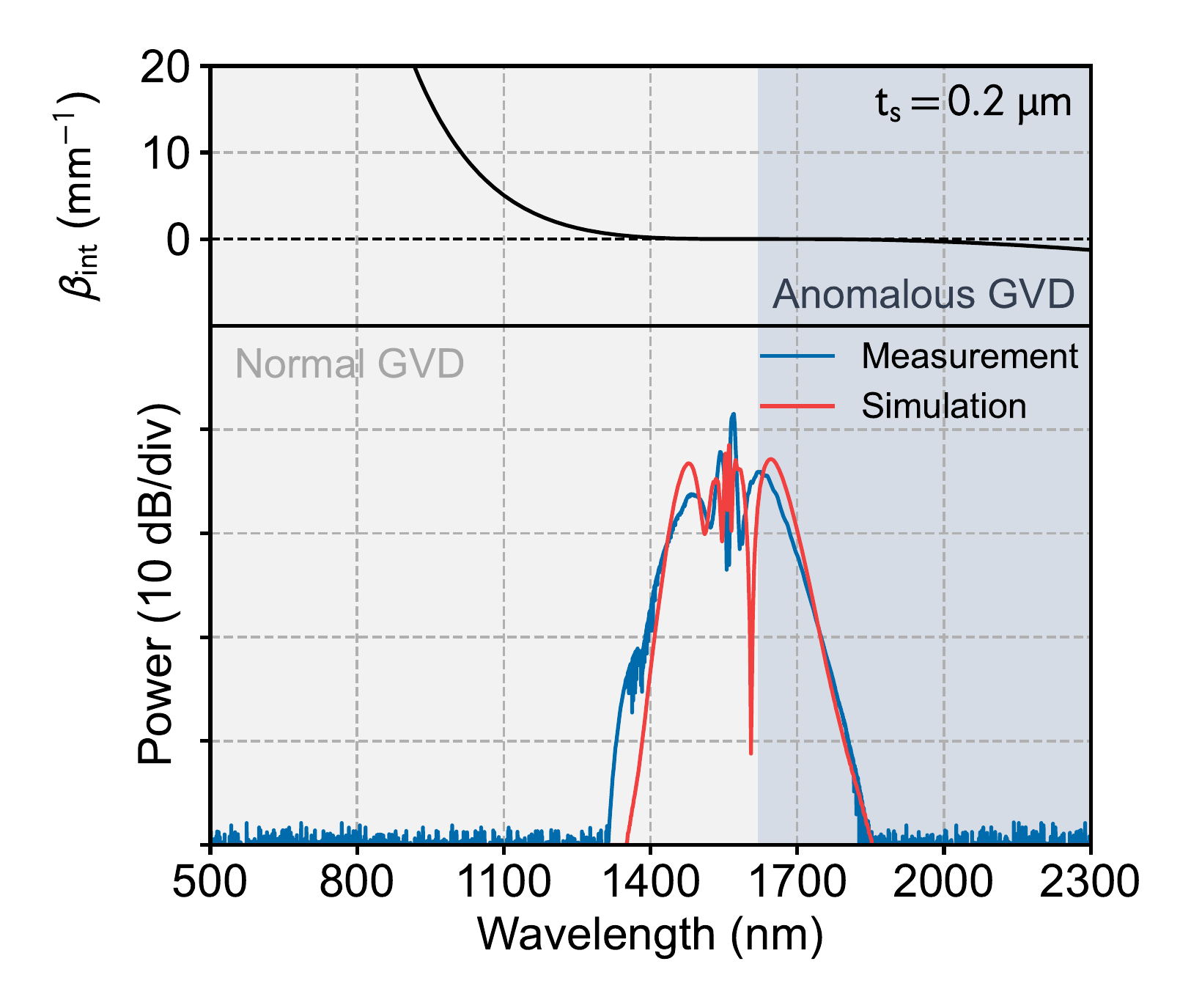}
	}
	\caption{Top: computed integrated dispersion. Bottom: comparison between the experimental (blue) and simulated (red) output spectra of the $\rm {t_s=0,\ 0.1\ \rm {and}\ 0.2\ \upmu m}$ waveguides.}
\end{figure}

The simulated pulse evolution within the three waveguide structures are shown in Figs. 5(a-c). The simulations were carried out for propagation distance up to 6 cm (approximately twice the actual length of the waveguides in the experiment) to fully illustrate the soliton dynamics and the DW generation process. In general, soliton dynamics are better understood from the perspective of feature lengths, where the input pulse can be regarded as a high order soliton with a soliton order $N = {(L_{\rm D}/L_{\rm NL})}^{0.5}$. Here, $L_{\rm D}$ is the dispersion length obtained as $L_{\rm D} = T_{\rm 0}^{\rm 2}/|\beta_{\rm 2}|$ where $T_{\rm 0}$ is the half width at $1/e$ intensity, and $L_{\rm NL}$ is the nonlinear length where $L_{\rm NL} = 1/\gamma P_{\rm 0}$. The soliton fission length $L_{\rm fiss} = L_{\rm D}/N$ and the modulation instability (MI) length $L_{\rm MI} = (2L_{\rm D}L_{\rm NL})^{0.5}$ are used to describe the lengths at which the corresponding effect will take place \cite{agrawal2019nonlinear}. Table 1 shows the calculated feature lengths for the $\rm {t_s=0\ \rm {and}\ 0.1\ \upmu m}$ waveguides.

\begin{table}[htbp]
	\begin{center}
		\small
		\caption{Feature lengths for $\rm {t_s=0\ \rm {and}\ 0.1\ \upmu m}$ waveguides.}
		\setlength{\tabcolsep}{4mm}{
			\begin{tabular}{cccccc} \toprule
				$\rm {t_s\ (\upmu m)}$ & $N$ & $L_{\rm D}\ {\rm (cm)}$ & $L_{\rm {NL}}\ {\rm (cm)}$ & $L_{\rm fiss}\ {\rm (cm)}$ & $L_{\rm MI}\ {\rm (cm)}$ \\ \midrule
				0                    & 8.1 & 16.3                    & 0.25                       & 2.02                       & 2.85                     \\
				0.1                  & 9.4 & 25.5                    & 0.29                       & 2.71                       & 3.83                     \\
				\bottomrule
			\end{tabular}}
	\end{center}
\end{table}

Due to the lower dispersion and nonlinearity, the soliton order $N$ of the $\rm {t_s=0.1\ \upmu m}$ waveguide is 9.4, which is slightly higher than 8.1 for the $\rm {t_s=0\ \upmu m}$ waveguide. For the same input pulses, they both have a very short $L_{\rm NL}$ (0.25 and 0.29 cm) and a long $L_{\rm D}$ (16.3 and 25.5 cm) in $\rm {t_s=0\ \rm {and}\ 0.1\ \upmu m}$ waveguides, respectively. It indicates that SPM is predominant during the onset of propagation, which can be observed in the range of 0 to $\sim$1.5 cm in Figs. 5(a) and 5(b). In the case where the positive and negative chirp compensates each other, the pulse is continuously compressed. Their shorter fission length (2.02 and 2.71 cm) compared to the MI length (2.85 and 3.83 cm) indicates that soliton fission plays a more important role in the broadening process compared to MI in both waveguides. Solitons are deterministically ejected before MI sidebands are fully-developed, without the formation of random localized structures in the temporal domain \cite{narhi2016real}. This avoids the observation of typical SC instabilities associated with broadband noise amplification through spontaneous MI processes \cite{solli2012fluctuations}. The longer $L_{\rm fiss}$ of 2.71 cm for the $\rm {t_s=0.1\ \upmu m}$ waveguide compared to 2.02 cm for the $\rm {t_s=0\ \upmu m}$ waveguide means that a longer propagation length is required to trigger soliton ejection, which is observed in Figs. 5(a) and 5(b). After soliton fission occurs at 1.6 and 2.2 cm (for the $\rm {t_s=0\ \rm {and}\ 0.1\ \upmu m}$ waveguides, respectively), some of the ejected solitons have ultrahigh peak power and ultrashort width, which makes $L_{\rm D}$ comparable with (or smaller than) $L_{\rm NL}$, so that dispersion starts to dominate and causes the individual pulses to break up and spread further apart. Meanwhile, the solitons shed a part of their energy to form DWs between 0.1 and 0.8 ps. The DW generation is stronger in the $\rm {t_s=0.1\ \upmu m}$ waveguide, which is due to the smaller distance between the pump and phase matching point. The temporal overlap of solitons with DW results in their mutual interaction through cross-phase modulation (XPM) and four wave mixing (FWM) \cite{agrawal2019nonlinear}. Such XPM and FWM yields the generation of various peaks merging together to form a broadband supercontinuum. For the case considering normal GVD region pumping, displayed in Fig. 5(c), the dispersion and nonlinearity are not mutually compensated. The pulse is progressively deformed, until eventually reaching optical wave breaking (pulse steepening and shock wave formation in Fig. 5(f) after $\sim$5 cm of propagation \cite{wetzel2016experimental}). In the meantime, SPM becomes progressively weaker and the corresponding SC spectral coverage is significantly limited.

\begin{figure}[htbp]
	\centering
	\subfigure[]{
		\includegraphics[width = 0.316\textwidth]{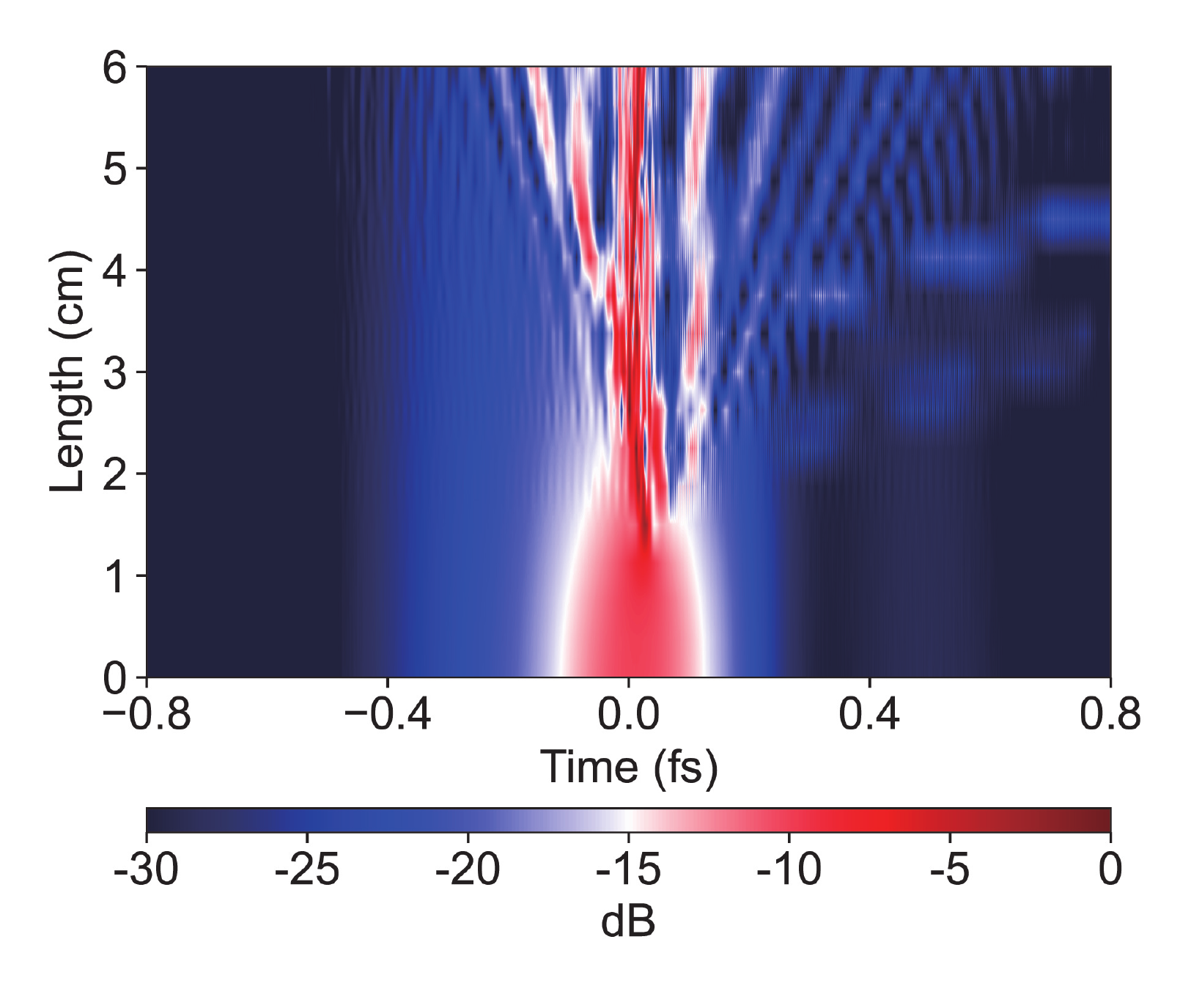}
	}
	\subfigure[]{
		\includegraphics[width = 0.316\textwidth]{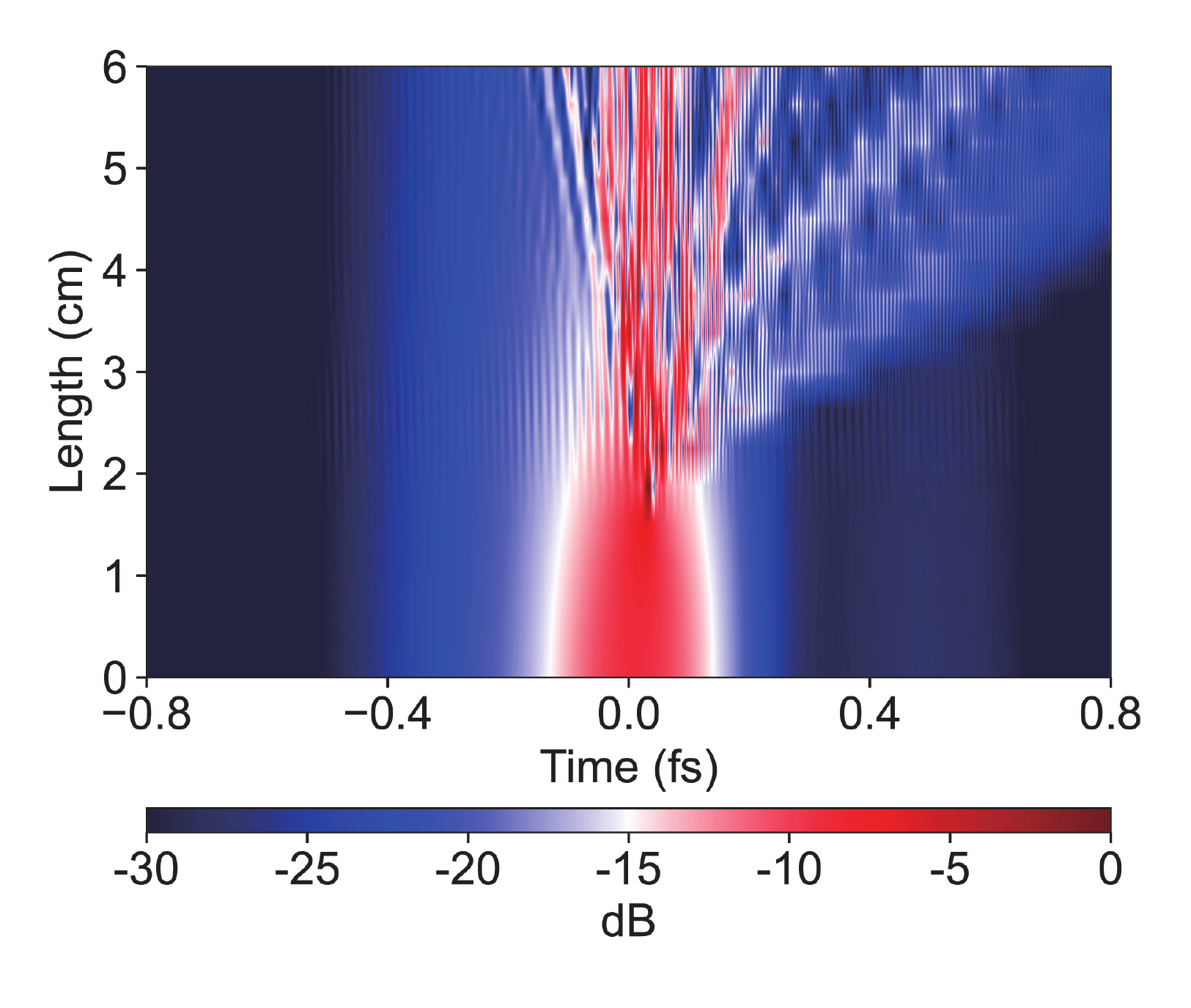}
	}
	\subfigure[]{
		\includegraphics[width = 0.316\textwidth]{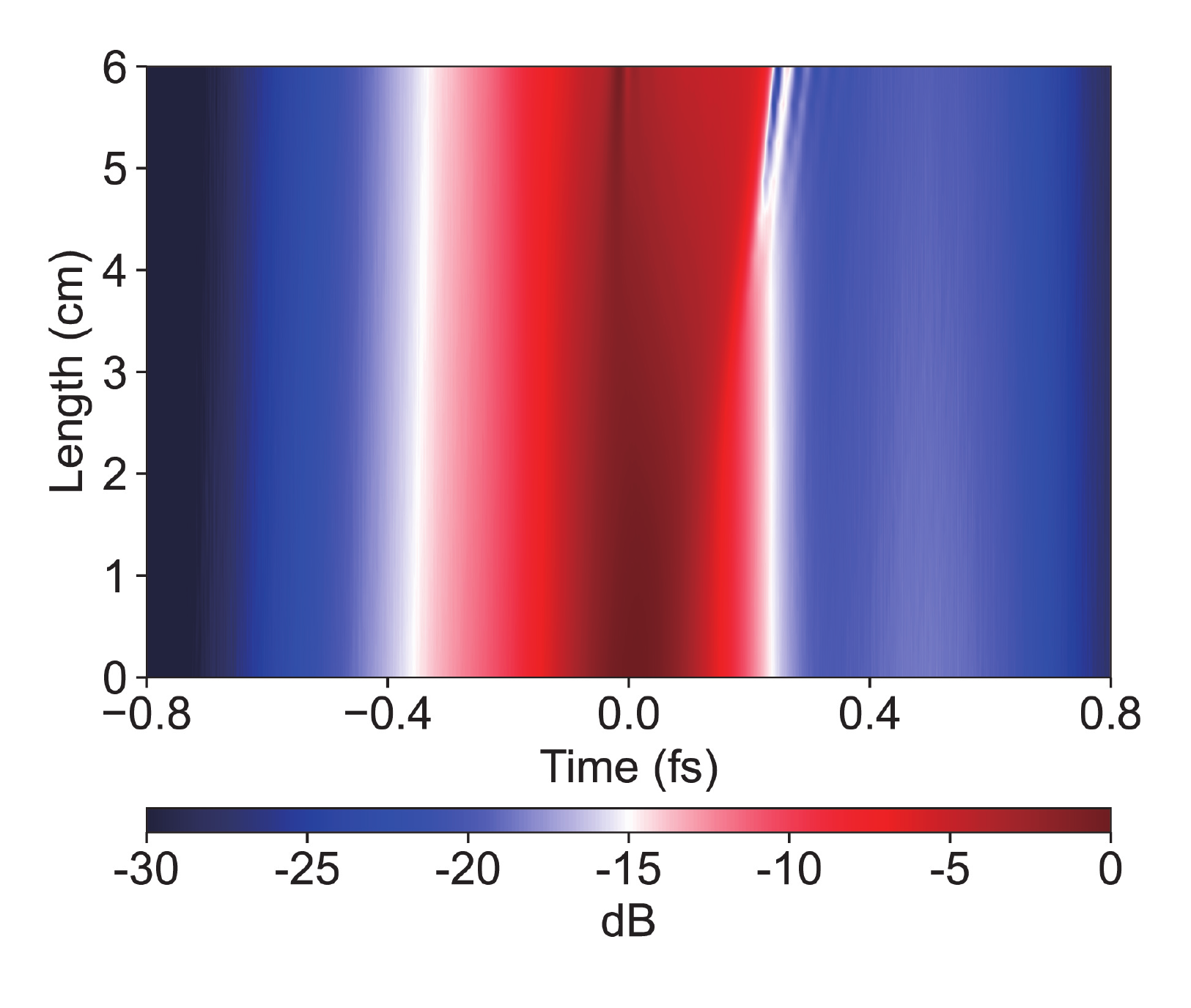}
	}
	\subfigure[]{
		\includegraphics[width = 0.31\textwidth]{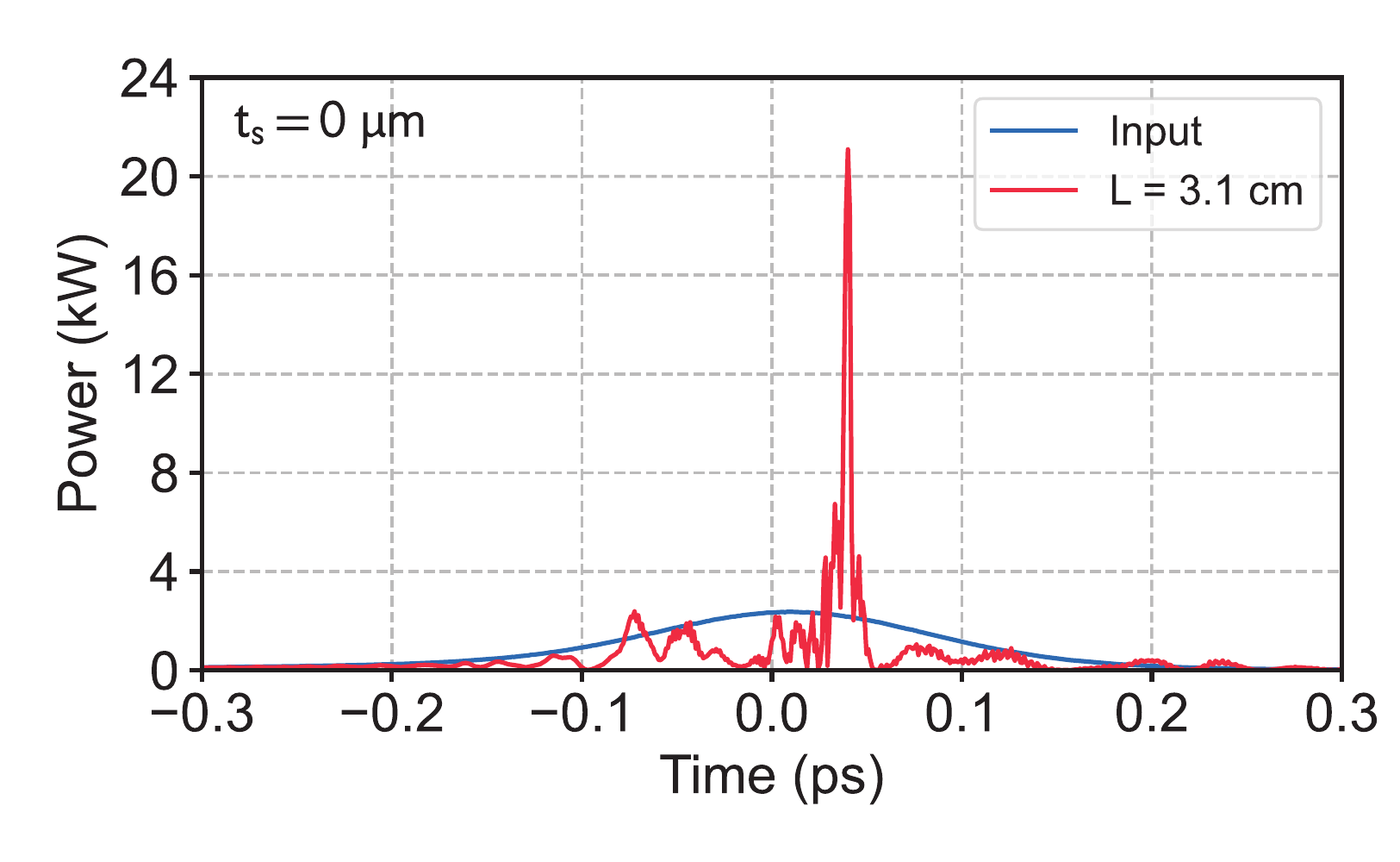}
	}
	\subfigure[]{
		\includegraphics[width = 0.31\textwidth]{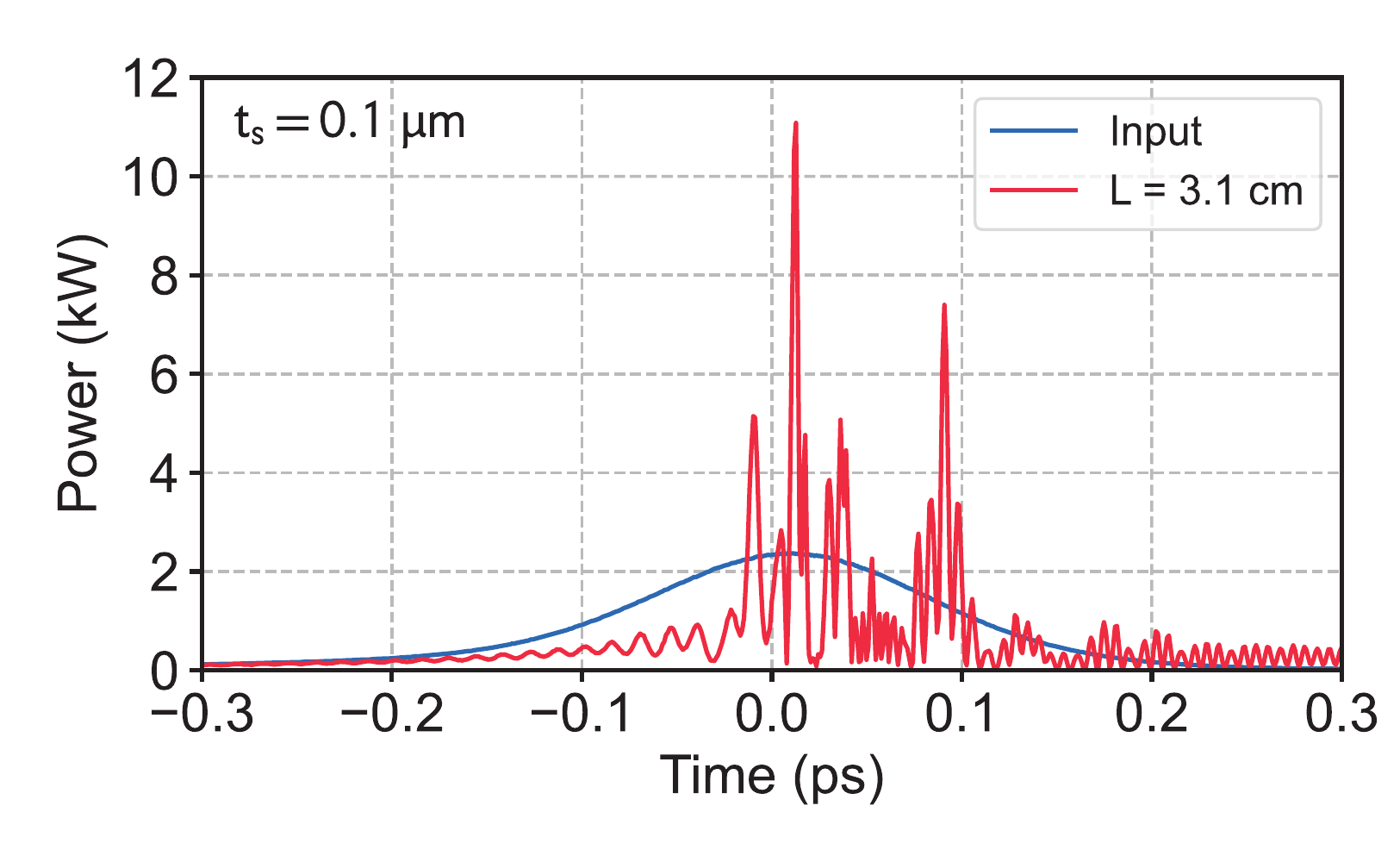}
	}
	\subfigure[]{
		\includegraphics[width = 0.31\textwidth]{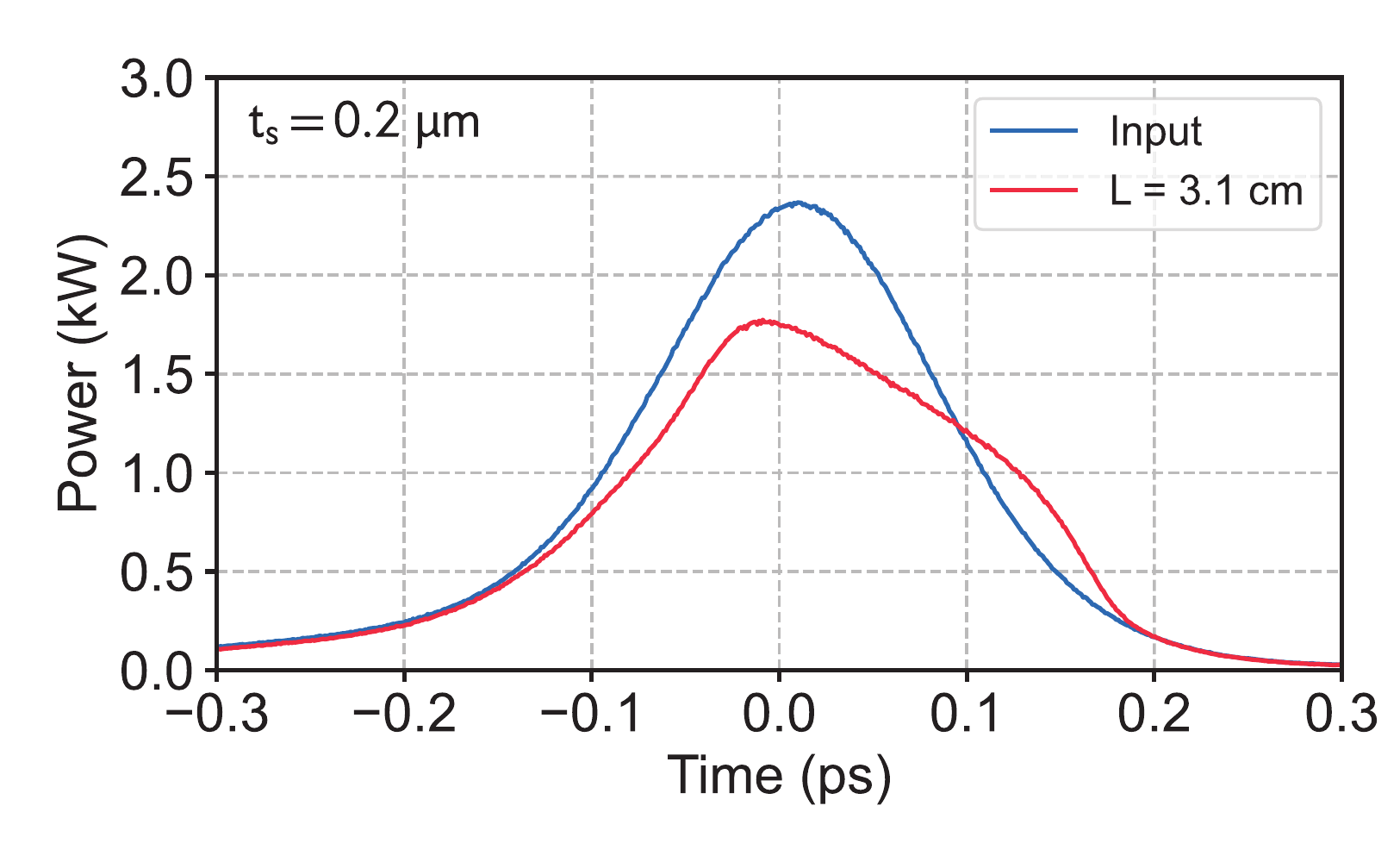}
	}
	\caption{Temporal evolution (a-c) of optical pulses in 6 cm long HDSG waveguides and corresponding temporal profiles at 3.1 cm (d-f). Numerical simulations considering waveguides with $\rm {t_s=0}$ (a,d), 0.1 (b,e) and 0.2 $\rm {\upmu m}$ (c,f).}
\end{figure}

Figures 5(d-f) show the simulated input (blue) and output (red) temporal waveforms at 3.1 cm waveguides corresponds to the spectrum in Figs. 4(a-c). For the $\rm {t_s=0\ \rm {and}\ 0.1\ \upmu m}$ waveguides in Figs 5(d) and 5(e), soliton fission happens within the 3.1 cm length of the waveguide and, at this distance, the generated solitons features a peak power of 8.3 and 4.65 times higher than the input pulse. All these solitons are much shorter than the input pulse while the shortest one being narrower by a factor of $2N-1$ \cite{wetzel2016experimental}. This results in the shortest soliton after fission is $\sim$10 fs. When pumping in the normal GVD region, as seen Fig. 5(c), the pulse loses its Gaussian shape without chirp compensation, the corresponding peak power being reduced by $\sim$25.5\%, and will continue to collapse during subsequent propagation.

\begin{figure}[htbp]
	\centering
	\subfigure[]{
		\includegraphics[width = 0.31\textwidth]{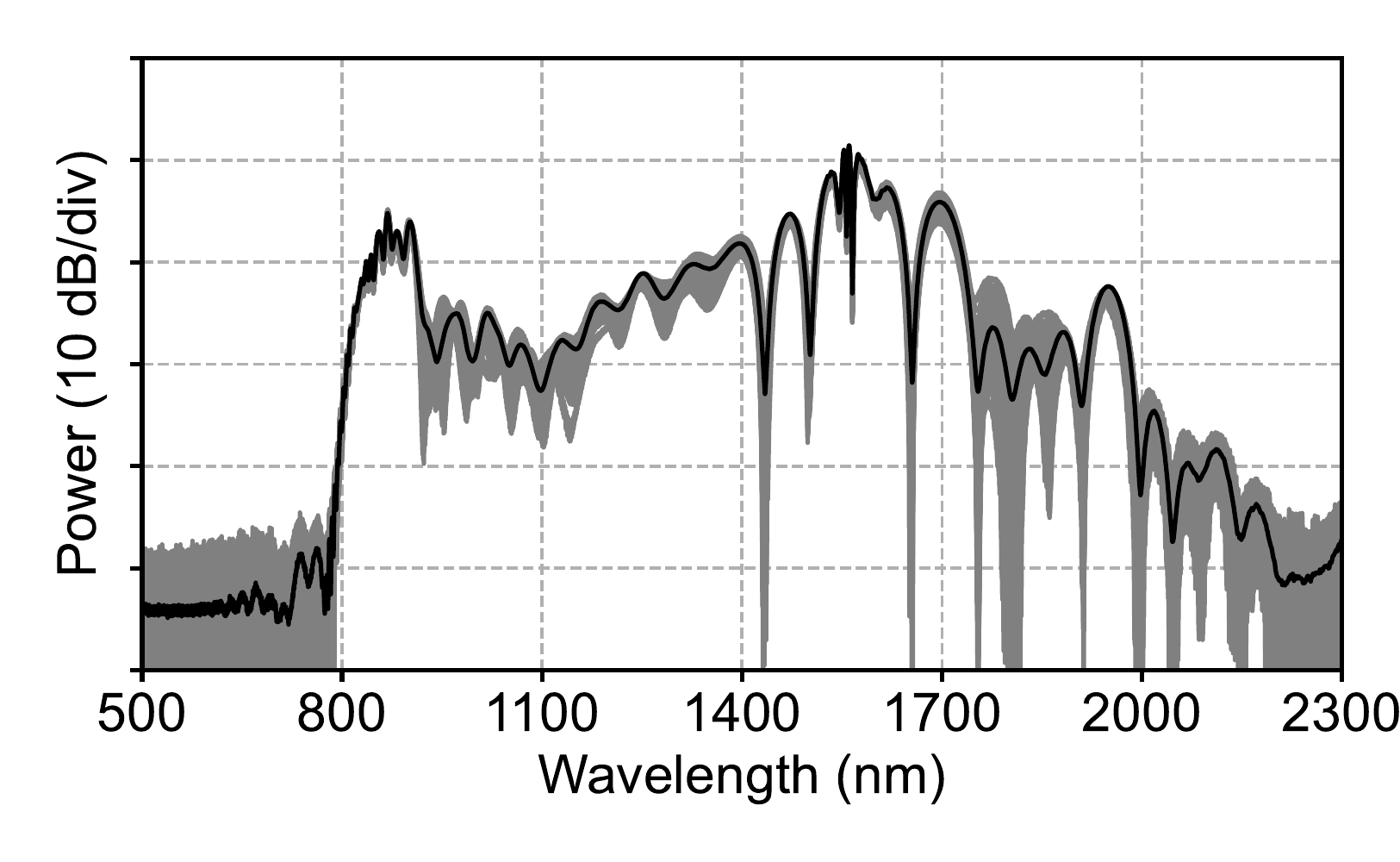}
	}
	\subfigure[]{
		\includegraphics[width = 0.31\textwidth]{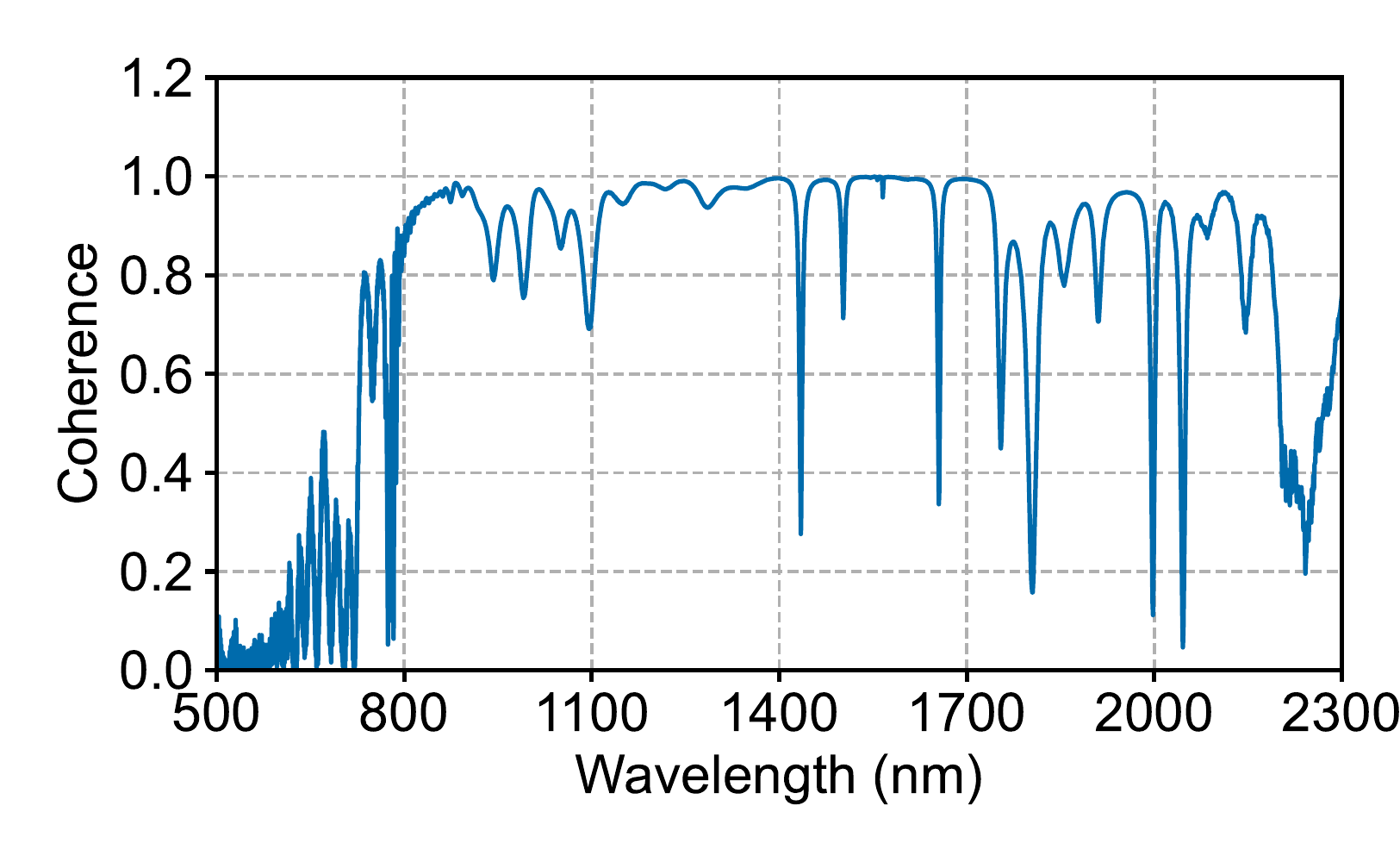}
	}
	\subfigure[]{
		\includegraphics[width = 0.31\textwidth]{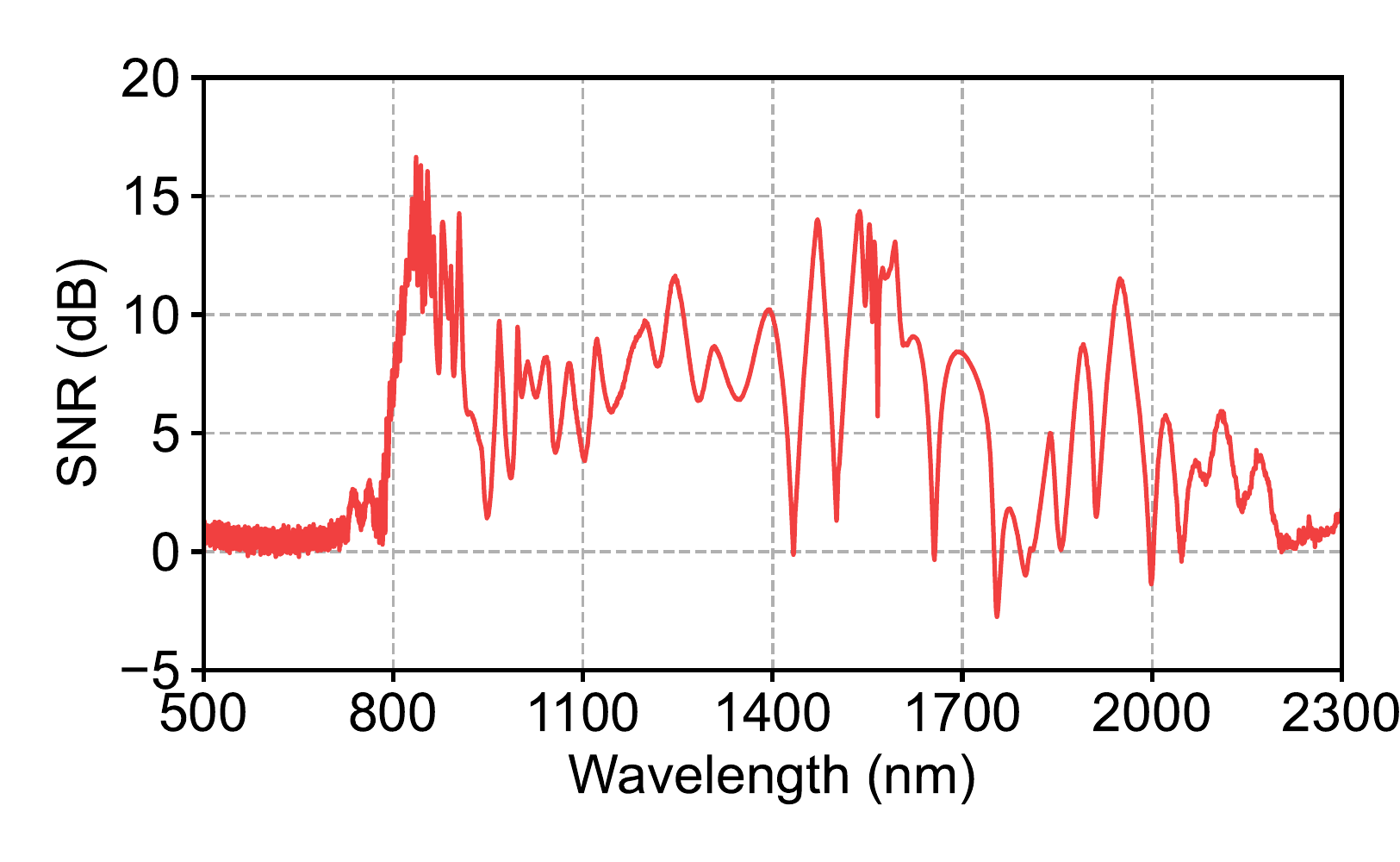}
	}
	\caption{Numerical assessment of SC stability: (a) Spectra from 256 Monte-Carlo simulations with random noise (gray) and corresponding average spectrum (black); (b) Calculated first-order coherence degree; (c) Calculated SNR.}
\end{figure}

To further verify the phase and intensity stability of the SC spectrum generated from the $\rm {t_s=0.1\ \upmu m}$ waveguide, we performed Monte-Carlo simulations including random vacuum noise (one photon per spectral mode with random phase \cite{dudley2006supercontinuum}) as well as intensity noise with a relative amplitude set at $10^{-4}$ of the pump pulse. Figure 6(a) shows 256 of those individual simulations with different random noise in gray and their average power in black. From those simulations, we retrieved both the SC spectral coherence and SNR. The first order coherence is calculated with the method in \cite{dudley2002coherence} while the SNR is evaluated as ${\rm SNR}(\omega) = \mu (\omega)/\sigma (\omega)$, where $\mu (\omega)$ is the average power and $\sigma (\omega)$ is standard deviation. The calculated coherence and SNR are shown in Figs. 6(b) and 6(c), respectively, they indicate the excellent stability and coherence degree across the whole SC bandwidth. Here, the coherence remains above 0.8 from 817 nm to 2183 nm, except for some wavelengths with minimal energy. The overall coherence is 0.92 which is comparable with the SC generated from silicon \cite{singh2018octave} and silicon germanium \cite{sinobad2019high} waveguides. From the SNR perspective, the highest SNR of $\sim$17 dB is found at the spectral location of DW formation, and otherwise fluctuates between -2.5 dB and 17 dB across the SC spectrum while the highest SNR is comparable with the SC generated from chalcogenide PCF \cite{huang2020mid}.

\section{Conclusion}

\noindent In summary, we have demonstrated and studied the dynamics of broadband SC generation in integrated CMOS-compatible and dispersion engineered HDSG waveguides. We have shown that the addition of a lower index oxide slot in the middle of the core of the conventional HDSG channel waveguide can be used to engineer the waveguide dispersion, where one can control the ZDW, phase matching points and the flatness of the dispersion curve. By introducing a 0.1 $\rm {\upmu m}$ oxide slot, a low and flat anomalous dispersion region can be achieved around the C-band, which leads to SC generation with a spectrum spanning over 1.5 octave, from 817 nm to 2183 nm, when pumped at 1560 nm, with high degree of coherence. We verified our experimental results through numerical simulations based on a GNLSE model, that showed a good match between DW locations and specific spectral features. With the ability to engineering its dispersion, the proposed HDSG slot waveguides thus offer a very promising realization of a fully integrated SC source that can find applications in frequency metrology and frequency synthesis, but also pave the way towards the integration of advanced on-chip functionalities within waveguide architectures with minimal dispersion, compatible with broadband and potentially reconfigurable optical signal processing.

\begin{backmatter}

	\vspace{10pt}
	\bmsection{Funding}
	\noindent This work was funded by the National Natural Science Foundation of China (62105291). This work was funded by the Shenzhen Fundamental Research Program (GXWD20201231165807007-20200827130534001). This work was funded by the European Research Council (ERC) under the European Union’s Horizon 2020 research and innovation programme under grant agreement No. 950618 (STREAMLINE project). Dr. Benjamin Wetzel further acknowledges the support of Région Nouvelle-Aquitaine (SCIR \& SPINAL projects).

	\bmsection{Disclosures}
	\noindent The authors declare no conflicts of interest.

	\bmsection{Data availability}
	\noindent Data underlying the results presented in this paper are not publicly available at this time but may be obtained from the authors upon reasonable request.

\end{backmatter}

\bibliography{Manuscript}

\end{document}